\documentclass[12pt]{article}
\pdfoutput=1
\usepackage[bookmarksnumbered=true,colorlinks=true,filecolor=blue,linkcolor=blue,urlcolor=blue,citecolor=red,linktocpage=true]{hyperref}
\usepackage{graphicx}
\usepackage{graphics}
\usepackage{dsfont}
\usepackage{epsfig}
\usepackage{amsmath,amssymb,amsthm,amscd}
\usepackage{bm}
\usepackage{tikz}
\usepackage{subcaption}

\def\ket{\rangle}
\def\bra{\langle}

\def\cD{\mathcal{D}}
\def\cL{\mathcal{L}}
\def\cO{\mathcal{O}}

\def\del{\partial}

\def\vol{\text{vol}}
\def\m{\text{m}}

\newcommand{\re}{\mathrm{e}}
\newcommand{\ri}{\mathrm{i}}
\newcommand{\rd}{\mathrm{d}}

\newcommand{\x}{\mathsf{x}}
\newcommand{\y}{\mathsf{y}}

\newcommand{\be}{\begin{equation}}
\newcommand{\ee}{\end{equation}}
\newcommand{\ba}{\begin{aligned}}
\newcommand{\ea}{\end{aligned}}
\numberwithin{equation}{section}

\begin{document}

\renewcommand{\thefootnote}{\fnsymbol{footnote}}
\setcounter{page}{0}
%%%%%%%%%%%%%%%%% Title page %%%%%%%%%%%%%%%%%%%%%%%%%%%%%%%%%
\thispagestyle{empty}
\begin{flushright}  
USTC-ICTS/PCFT-20-17
 \end{flushright} %paper number

\vskip1.2cm
\begin{center}
{\Large {\bf Quantum Periods and Spectra in  \\
\vskip 0.2 cm 
Dimer Models and Calabi-Yau Geometries}}

 \vskip1.2cm
{\large 
 %\bf
  Min-xin Huang\footnote{\href{mailto: minxin@ustc.edu.cn}{\tt minxin@ustc.edu.cn}},
  Yuji Sugimoto\footnote{\href{mailto: sugimoto@ustc.edu.cn}{\tt sugimoto@ustc.edu.cn}},
   Xin Wang\footnote{\href{mailto: wxin@mpim-bonn.mpg.de}{\tt wxin@mpim-bonn.mpg.de}} 
}
%name address

\vskip1.2cm

$^{*} $$^{\dagger}$
Interdisciplinary Center for Theoretical Study,  \\ \vskip 0.1cm  University of Science and Technology of China,  Hefei, Anhui 230026, China
 \\ \vskip 0.3 cm
$^{*} $$^{\dagger}$ 
Peng Huanwu Center for Fundamental Theory,  \\ \vskip 0.1cm  Hefei, Anhui 230026, China

\vskip.3cm

$^{\ddagger}$
 Bethe Center for Theoretical Physics, Universit\"at Bonn, D-53115, Bonn, Germany
\vskip.3cm
$^{\ddagger}$
Max Planck Institute for Mathematics, Vivatsgasse 7, D-53111 Bonn, Germany

\end{center}

\vskip1cm
\begin{abstract} %abstract
We study a class of quantum integrable systems derived from dimer graphs and also described by local toric Calabi-Yau geometries with higher genus mirror curves, generalizing some previous works on genus one mirror curves. We compute the spectra of the quantum systems both by standard perturbation method and by Bohr-Sommerfeld method with quantum periods as the phase volumes. In this way, we obtain some exact analytic results for the classical and quantum periods of the Calabi-Yau geometries. We also determine the differential operators of the quantum periods and compute the topological string free energy in Nekrasov-Shatashvili (NS) limit. The results agree with calculations from other methods such as the topological vertex.

\end{abstract}

%%%%%%%%%%%%%%%%%%%%%%%%%%%%%%%%%%%%%%%%%%%%%%%%%%%%%%%%
\renewcommand{\thefootnote}{\arabic{footnote}}
\setcounter{footnote}{0}

\vfill\eject

\tableofcontents

\baselineskip=16pt

\section{Introduction and Summary}

The developments of various prosperous topics in mathematics and physics often intersect with each other. Topological string theory on Calabi-Yau manifolds has been a fruitful branch of superstring theories that encompass many recurring themes in mathematical physics, see e.g. \cite{ADKMV}. In the seminal work \cite{NS1}, Nekrasov and Shatashvili (NS) proposed a connection between the partition function of Seiberg-Witten gauge theory on $\Omega$ background and certain quantum integrable systems. In the NS limit, we set one of the two $\Omega$ deformation parameters to vanish and identify the other as the Planck constant of the quantum system. This relation can be uplifted to five dimensions, where the partition functions are computed by refined topological string theory on corresponding Calabi-Yau spaces. The topological string free energy in the NS limit can be viewed as a quantum deformation of the prepotential and is computed similarly by promoting the periods of the Calabi-Yau geometries to quantum periods \cite{MM, ACDKV, Huang1, HKRS}. More examples in Seiberg-Witten theories can be found in \cite{Mironov:2009dv, Bourgine:2012gy, Fucito:2012xc, Billo:2013fi, Ito:2017iba}. The quantization conditions of the quantum system are formulated as the Bohr-Sommerfeld quantization conditions where the phase volumes are computed by quantum periods. In the five-dimensional case, the quantum systems are often known as relativistic models due to the exponential kinetic and potential terms in the Hamiltonians from quantizing the mirror curves of the local Calabi-Yau spaces. Inspired by earlier works \cite{Hatsuda:2013oxa, Kallen:2013qla, HW}, some novel non-perturbative contributions to the quantization conditions are conjectured in \cite{Grassi:2014, Wang:2015}. Various aspects of the quantization conditions, including complex value Planck constant, resurgence, wave functions, etc are further explored in, e.g., \cite{Kashani-Poor:2016edc, CMS, Grassi:2017qee, Sciarappa:2017hds, Zakany:2017txl}. The non-perturbative parts of the two types of exact quantization conditions in \cite{Grassi:2014, Wang:2015} are related by certain constrains on the BPS invariants known as the blowup equations \cite{Sun:2016obh, Grassi:2016nnt}. The blowup equations originally come from studies of Seiberg-Witten gauge theories \cite{Gottsche:2006bm} (see also \cite{Keller:2012da, Kim:2019uqw}), but have now become a very effective tool for computing topological string amplitudes on various Calabi-Yau manifolds \cite{Huang:2017mis, Gu:2018gmy, Gu:2019dan, Gu:2019pqj, Gu:2020fem}. The exact quantization conditions have also been applied to related condensed matter systems, e.g., in \cite{Hatsuda:2016mdw, Hatsuda:2017zwn, Duan:2018dvj, Hatsuda:2020ocr}. 

Most examples of the early studies focus on geometries with mirror curves of genus one. The quantum periods and quantization conditions for quantum systems corresponding to mirror curves of the higher genus were subsequently considered in, e.g., \cite{Sun:2016obh, CGM1, Codesido:2016ixn, Grassi:2018bci, Fischbach:2018yiu}. A particularly interesting class of quantum integrable systems can be constructed by dimer models on torus \cite{MR3185352}, and the quantization conditions are studied in \cite{HM1, SHM1}. The dimer models in this paper also correspond to local toric Calabi-Yau geometries and the mirror curves are encoded in the data of the bipartite dimer graphs. Some of Calabi-Yau spaces geometrically engineer 5d supersymmetric gauge theories, which are uplifts of the 4d $SU(N)$ Seiberg-Witten theories considered in \cite{Mironov:2009dv}. There are a number of commuting Hamiltonians, and the multiple quantization conditions can be similarly derived from topological string free energy in the NS limit on the corresponding Calabi-Yau spaces. The studies in \cite{HM1, SHM1} mostly focus on numerical tests of the non-perturbative quantization conditions. However, in order to have a more insightful understanding of the interconnections between various subjects here, it is better to have some analytical results. In this paper, we develop some analytic approaches to the problem, though mostly focusing on the perturbative aspects.

The paper is organized as follows. In Section \ref{sec:dimermodel} we review the constructions of dimer models, and derive Hamiltonians of the quantum integrable systems based on previous literature. We shall study some examples with genus two mirror curves and correspondingly two commuting dynamical Hamiltonians. In Section \ref{secperturbative} we study the perturbative quantum spectra of the Hamiltonians around minimal points of the phase space. A useful technical ingredient is the symplectic transformations of the quantum canonical coordinates, which are necessary to determine the energy eigenvalues of the quadratic terms. We find the symplectic transformations for the examples with simple classical minima, and further calculate the higher-order spectra with standard perturbation methods in quantum mechanics. In Section \ref{sectiontopo} we systematically compute the classical/quantum periods and topological string free energies for the Calabi-Yau geometries, summarizing the results in previous literature. We then compute the differential operators which exactly determine quantum corrections to classical periods, generalizing earlier works \cite{Huang1, HKRS} to the situation of higher genus mirror curves. Similarly, the topological free energy in the NS limit is determined by the quantum periods, and we show that this agrees with results from, e.g., method of the topological vertex. An interesting feature is that the differential operators are the same for different cycles of the higher genus mirror curves. Following earlier works \cite{HW}, we perform some satisfying tests of our calculations by comparing the quantum spectra from direct perturbation and Bohr-Sommerfeld quantization conditions using quantum periods as phase volumes. These exercises provide some exact analytic results for the classical and quantum periods of the Calabi-Yau spaces, which are difficult to directly obtain.

%%%%%%%%%%%%%%%%%%%%%%%%%%%%%%%%%%%%%%%%%%%%%%%%%

\section{Dimer models and integrable systems}\label{sec:dimermodel}
%In this section, we review the connection between dimer models and integrable systems in \cite{Eager:2011dp}

In \cite{MR3185352}, the authors proposed an infinite class of cluster integrable systems.\footnote{For $A$ type Toda systems, \cite{Williams:2012fz}\cite{Marshakov:2012kv} have an equivalent but different description.} The most interesting ones among them are the cluster integrable systems for the dimer models on a torus. The dimer model is the study of the set of perfect matching of a graph, where the perfect matching is a subset of edges that covers each vertex exactly once. For a bipartite graph, the vertices are divided into two sets, the black set, and the white set. Every edge connects a white vertex to a black vertex. For a more detailed introduction to dimer models, see \cite{MR2198850}. 

The dimer model can be connected to a toric diagram by Kasteleyn matrix $K(X,Y)$ \cite{MR2198850}, which is the weighted adjacency matrix of the graph. The determinant of the Kasteleyn matrix happens to be the mirror curve of the corresponding toric Calabi-Yau three-fold \cite{Hanany:2005ve}\cite{Franco:2005rj}, the adjacency matrix can be computed as follows:
\begin{itemize}
\item Multiply each edge weight of the graph a sign $\pm 1$, so that around every face, the product of the edge weights over edges bounding the face is
\be
\text{sgn}( \prod_i e_i)= 
\begin{cases}
  +1,& \text{if (\# edges)}=2\mod 4 \\
  -1,       & \text{if (\# edges)}=0\mod 4
\end{cases}
\ee
\item Construct two loops $\gamma_X,\gamma_Y$ along the two cycles of the torus, we draw them as red dash lines in the diagram. 
\item Fix an orientation, from black to white, as the positive orientation.
\item Multiply each edge with a factor $X$ or $Y$, if the loop $\gamma_X$ or $\gamma_Y$ get through the edge with positive orientation. Multiply each edge with a factor $1/X$ or $1/Y$, if the loop $\gamma_X$ or $\gamma_Y$ get through the edge with positive orientation. 
\end{itemize}
Then the Kasteleyn matrix is a matrix with rows labeled by black vertices and columns labeled by white vertices, with the entry as the weight between the connected black and white vertices. The entry is 0 if two points are not connected. In this paper, we are interested in $Y^{p,q}$ system, which is originally introduced in \cite{Benvenuti:2004dy}, the determinant of the Kasteleyn matrix has the form
\be
Y+\frac{X^q}{Y}+X^{p+2}+\cdots+X+1=0.
\ee

\begin{figure}
\center
\begin{tikzpicture}
\draw[->,red,thick] (0.5 ,-0.1) -- (2 ,-0.1)-- (2+0.65,-0.65-0.1);
\draw[->,blue,thick] (0.5 ,0.1) -- (2 ,0.1)--(2+0.65,0.65+0.1);
\draw[fill=white] (2 cm, 0 cm) circle(.1cm) node[below=3pt]{$v$};
\draw (2+0.65,0.65+0.1) node[right=3pt]{$\omega_1$};
\draw (2+0.65,-0.65-0.1) node[right=3pt]{$\omega_2$};
\draw (0,-1.35-0.1) node[right=3pt]{(a)\quad$\delta_v(\omega_1,\omega_2)=-\frac{1}{2}$};
\end{tikzpicture}
\hspace*{1 cm}
\begin{tikzpicture}
\draw[->,red,thick] (0.5 ,-0.1) -- (2 ,-0.1)-- (2+0.65,-0.65-0.1);
\draw[<-,blue,thick] (0.5 ,0.1) -- (2 ,0.1)--(2+0.65,0.65+0.1);
\draw[fill=white] (2 cm, 0 cm) circle(.1cm) node[below=3pt]{$v$};
\draw (2+0.65,0.65+0.1) node[right=3pt]{$\omega_1$};
\draw (2+0.65,-0.65-0.1) node[right=3pt]{$\omega_2$};
\draw (0,-1.35-0.1) node[right=3pt]{(b)\quad$\delta_v(\omega_1,\omega_2)=\frac{1}{2}$};
\end{tikzpicture}
\hspace*{1 cm}
\begin{tikzpicture}
\draw[->,blue,thick] (0.5 ,-0.1) -- (2 ,-0.1)-- (2+0.65,-0.65-0.1);
\draw[->,red,thick] (0.5 ,0.1) -- (2 ,0.1)--(2+0.65,0.65+0.1);
\draw[fill=white] (2 cm, 0 cm) circle(.1cm) node[below=3pt]{$v$};
\draw (2+0.65,0.65+0.1) node[right=3pt]{$\omega_2$};
\draw (2+0.65,-0.65-0.1) node[right=3pt]{$\omega_1$};
\draw (0,-1.35-0.1) node[right=3pt]{(c)\quad$\delta_v(\omega_1,\omega_2)=\frac{1}{2}$};
\end{tikzpicture}
\caption{An illustration of $\delta_v(\omega_1,\omega_2)$. If $\omega_1$ and $\omega_2$ are in the counterclockwise order, and with the same direction, $\delta_v(\omega_1,\omega_2)=\frac{1}{2}$ as in (c). Any change in the clockwise order or direction gives an extra sign, e.g., (a)(b). The arrows represent the orientations of the loops $\omega_i$. }
\label{fig:deltav}
\end{figure}
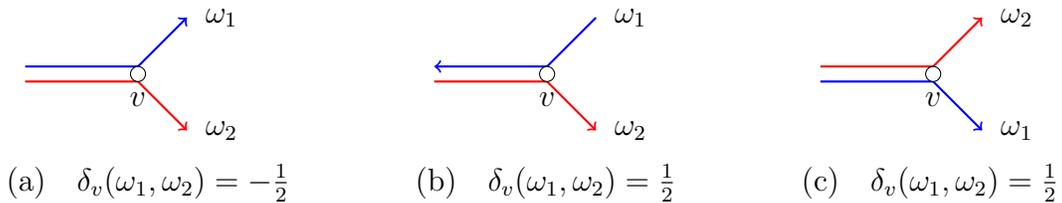

Following \cite{MR3185352}\cite{Eager:2011dp}, the commutation relations and the Hamiltonians of the cluster integrable systems can be read from the loops of the graph. %We first fix a perfect matching, as the reference perfect matching, then the loops can be defined from the difference from the reference perfect matching and another perfection matching. 
Let $\omega_i$ be the oriented loops on the graph, the Poisson bracket between cycles are defined as
\be
\{\omega_i,\omega_j\}=\epsilon_{\omega_i,\omega_j} \omega_i\omega_j,
\ee
where
\be
\epsilon_{\omega_i,\omega_j} :=\sum_v \text{sgn}(v) \delta_v(\omega_i,\omega_j).
\ee
Here $\text{sgn}(v)=1$ for the white vertex $v$, and $-1$ for the black vertex. $\delta_v$ is a skew symmetric bilinear form with $\delta_v(\omega_i,\omega_j)=-\delta_v(\omega_j,\omega_i)=-\delta_v(-\omega_i,\omega_j)\in \frac{1}{2}\mathbb{Z}$, as illustrated in Figure~\ref{fig:deltav}. Though more general vertex is possible, for our examples of dimer models we will only encounter cubic vertices. 

\begin{figure}
\center
\begin{tikzpicture}
%white nodes
\foreach \position in {(0,0),(0,1),(0,2),(0,3),(1.5,-0.5),(1.5,0.5),(1.5,1.5),(1.5,2.5),(3,0),(3,1),(3,2)}
	\draw[fill=white] \position circle(.05cm);
%black nodes
\foreach \position in {(1,0),(1,1),(1,2),(1,3),(2.5,-0.5),(2.5,0.5),(2.5,1.5),(2.5,2.5),(-0.5,0.5),(-0.5,1.5),(-0.5,2.5)}
	\draw[fill=black] \position circle(.05cm);
%red lines
\foreach \position in {(0,0), (0,1), (0,2), (0,3),(1.5,-0.5),(1.5,0.5),(1.5,1.5),(1.5,2.5)}
  \draw[red, line width=0.06cm] \position+(0.05,0) -- ++(0.95,0);
%blue lines
\foreach \position in {(2.5,-0.5),(2.5,0.5),(2.5,1.5),(-0.5,0.5),(-0.5,1.5),(-0.5,2.5)}
  \draw[] \position +(0.04,0.04)-- ++(0.46,0.46);
\foreach \position in {(1,0),(1,1),(1,2),(1,3)}
  \draw[] \position +(0.04,-0.04)-- ++(0.46,-0.46);
%green lines  
\foreach \position in {(2.5,0.5),(2.5,1.5),(2.5,2.5),(-0.5,0.5),(-0.5,1.5),(-0.5,2.5)}
  \draw[] \position+(0.04,-0.04) -- ++(0.46,-0.46);
\foreach \position in {(1,0),(1,1),(1,2)}
  \draw[] \position+(0.04,0.04) -- ++(0.46,0.46);   
\end{tikzpicture}
\hspace*{1 cm}
\begin{tikzpicture}
%white nodes
\foreach \position in {(0,0),(0,1),(0,2),(0,3),(1.5,-0.5),(1.5,0.5),(1.5,1.5),(1.5,2.5),(3,0),(3,1),(3,2)}
	\draw[fill=white] \position circle(.05cm);
%black nodes
\foreach \position in {(1,0),(1,1),(1,2),(1,3),(2.5,-0.5),(2.5,0.5),(2.5,1.5),(2.5,2.5),(-0.5,0.5),(-0.5,1.5),(-0.5,2.5)}
	\draw[fill=black] \position circle(.05cm);
%red lines
\foreach \position in {(0,0), (0,1), (0,2), (0,3),(1.5,-0.5),(1.5,0.5),(1.5,1.5),(1.5,2.5)}
  \draw[red, line width=0.06cm] \position+(0.05,0) -- ++(0.95,0);
%blue lines
\foreach \position in {(2.5,-0.5),(2.5,0.5),(2.5,1.5),(-0.5,0.5),(-0.5,1.5),(-0.5,2.5)}
  \draw[blue,line width=0.06cm] \position +(0.04,0.04)-- ++(0.46,0.46);
\foreach \position in {(1,0),(1,1),(1,2),(1,3)}
  \draw[blue,line width=0.06cm] \position +(0.04,-0.04)-- ++(0.46,-0.46);
%green lines  
\foreach \position in {(2.5,0.5),(2.5,1.5),(2.5,2.5),(-0.5,0.5),(-0.5,1.5),(-0.5,2.5)}
  \draw[] \position+(0.04,-0.04) -- ++(0.46,-0.46);
\foreach \position in {(1,0),(1,1),(1,2)}
  \draw[] \position+(0.04,0.04) -- ++(0.46,0.46);   
\end{tikzpicture}
\hspace*{1 cm}
\begin{tikzpicture}
%white nodes
\foreach \position in {(0,0),(0,1),(0,2),(0,3),(1.5,-0.5),(1.5,0.5),(1.5,1.5),(1.5,2.5),(3,0),(3,1),(3,2)}
	\draw[fill=white] \position circle(.05cm);
%black nodes
\foreach \position in {(1,0),(1,1),(1,2),(1,3),(2.5,-0.5),(2.5,0.5),(2.5,1.5),(2.5,2.5),(-0.5,0.5),(-0.5,1.5),(-0.5,2.5)}
	\draw[fill=black] \position circle(.05cm);
%red lines
\foreach \position in {(0,0), (0,1), (0,2), (0,3),(1.5,-0.5),(1.5,0.5),(1.5,1.5),(1.5,2.5)}
  \draw[red, line width=0.06cm] \position+(0.05,0) -- ++(0.95,0);
%blue lines
\foreach \position in {(2.5,-0.5),(2.5,0.5),(2.5,1.5),(-0.5,0.5),(-0.5,1.5),(-0.5,2.5)}
  \draw[] \position +(0.04,0.04)-- ++(0.46,0.46);
\foreach \position in {(1,0),(1,1),(1,2),(1,3)}
  \draw[] \position +(0.04,-0.04)-- ++(0.46,-0.46);
%green lines  
\foreach \position in {(2.5,0.5),(2.5,1.5),(2.5,2.5),(-0.5,0.5),(-0.5,1.5),(-0.5,2.5)}
  \draw[green,line width=0.06cm] \position+(0.04,-0.04) -- ++(0.46,-0.46);
\foreach \position in {(1,0),(1,1),(1,2)}
  \draw[green,line width=0.06cm] \position+(0.04,0.04) -- ++(0.46,0.46);   
\end{tikzpicture}
\caption{An illustration of a reference perfect matching (red) and loops defined from two other perfect matchings (blue and green).}
\label{fig:deltav2}
\end{figure}
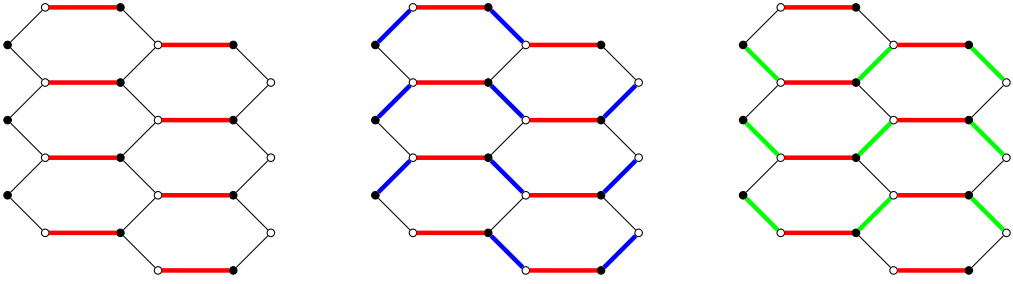

As described in \cite{MR3185352}, the subtraction of two different perfect matchings\footnote{More precisely, it is a subtraction of 1-chains defined from two perfect matchings. The 1-chain $[M]$ is a sum of oriented edges in the perfect matching $M$.} forms a cycle. To construct the basis $t_i$ of all the loops, we can first fix an arbitrary perfect matching as the reference perfect matching, and select the independent basis from the subtraction of other perfect matchings. For example, Figure \ref{fig:deltav2} is the unit of brane tiling for the $Y^{3,3}$ system, we chose the collection of red edges as the reference perfect matching $M_1$. Denote the collection of blue and green edges as the perfect matching $M_2$ and $M_3$ respectively, then $[M_2]-[M_1]$ and $[M_3]-[M_1]$ give the independent 1-loops $t_1,\cdots,t_6$ in (\ref{Y33cycle}). Given the coordinate basis, the Hamiltonians $H_n$ are defined from the sum of all $n$-loops, where the $n$-loop is the product of $n$ disjoint 1-loops in a coordinate expression. 

\subsection{Examples}
\begin{figure}\label{Y3qtiling}
\hspace{-1.2 cm}
\begin{subfigure}{0.6\textwidth}
\includegraphics[width=\textwidth]{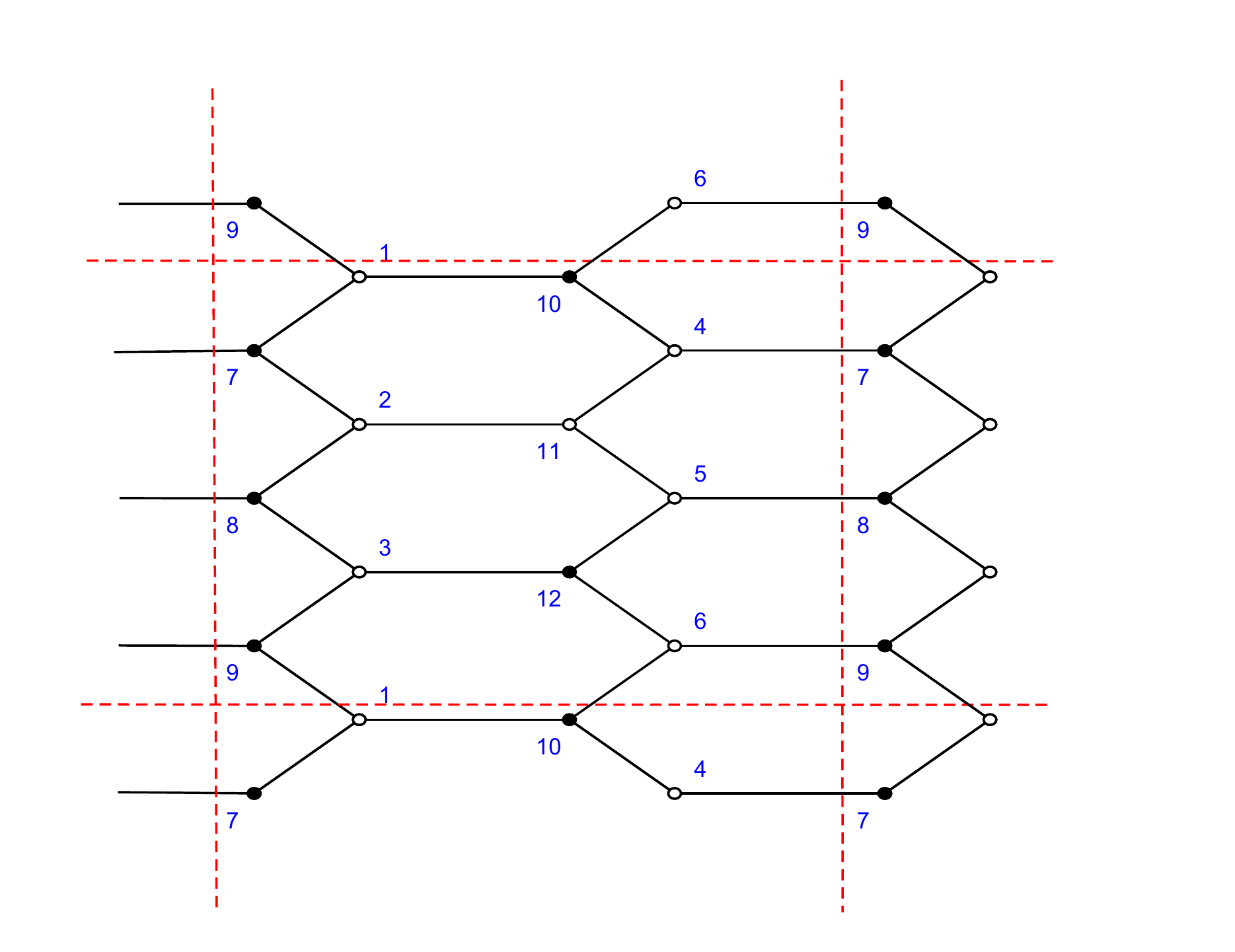}
\caption{$Y^{3,3}$}\label{fig:Y33}
\end{subfigure}
\begin{subfigure}{0.6\textwidth}
\includegraphics[width=\textwidth]{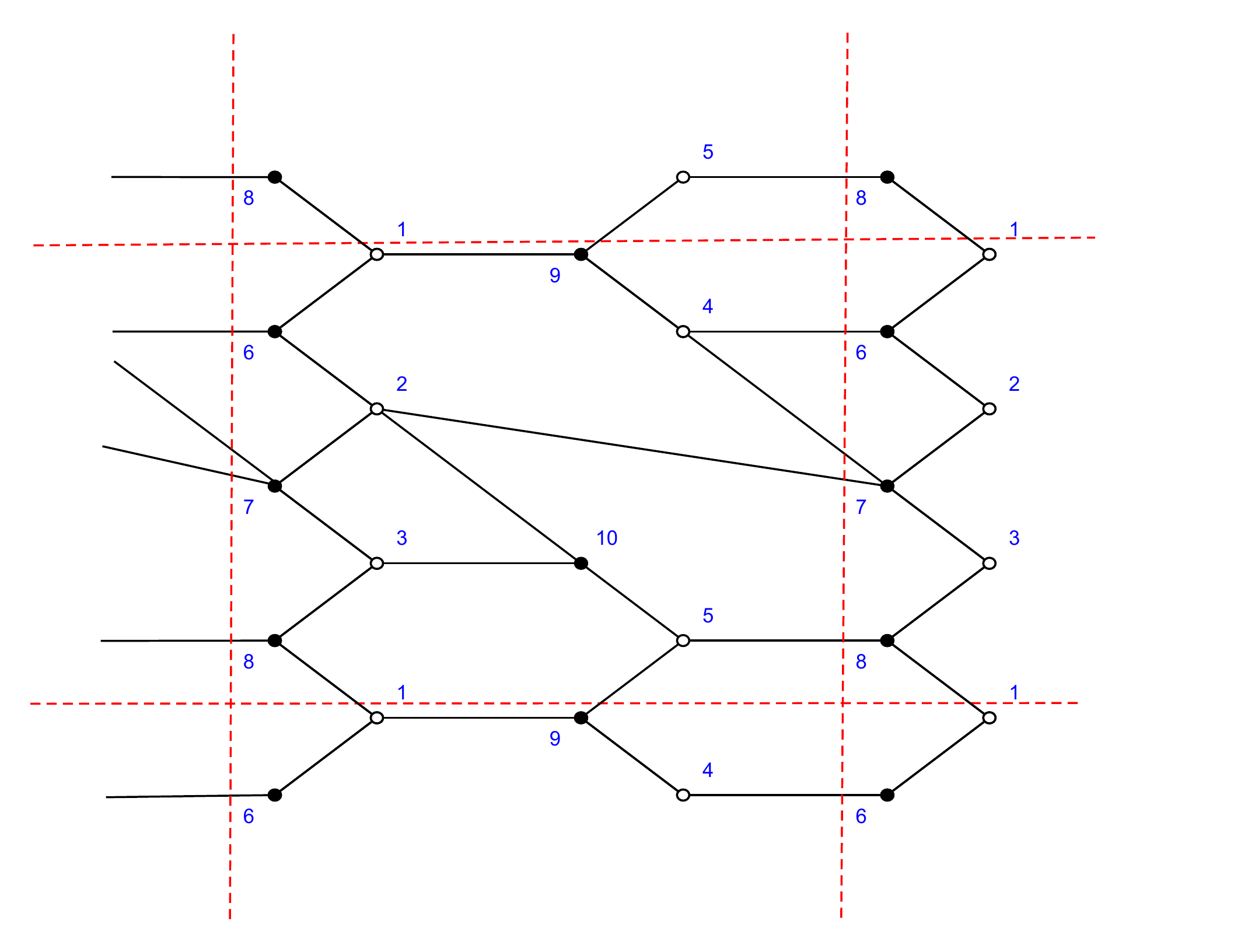}
\caption{$Y^{3,2}$}\label{fig:Y32}
\end{subfigure}

\hspace{-1.2 cm}
\begin{subfigure}{0.6\textwidth}
\includegraphics[width=\textwidth]{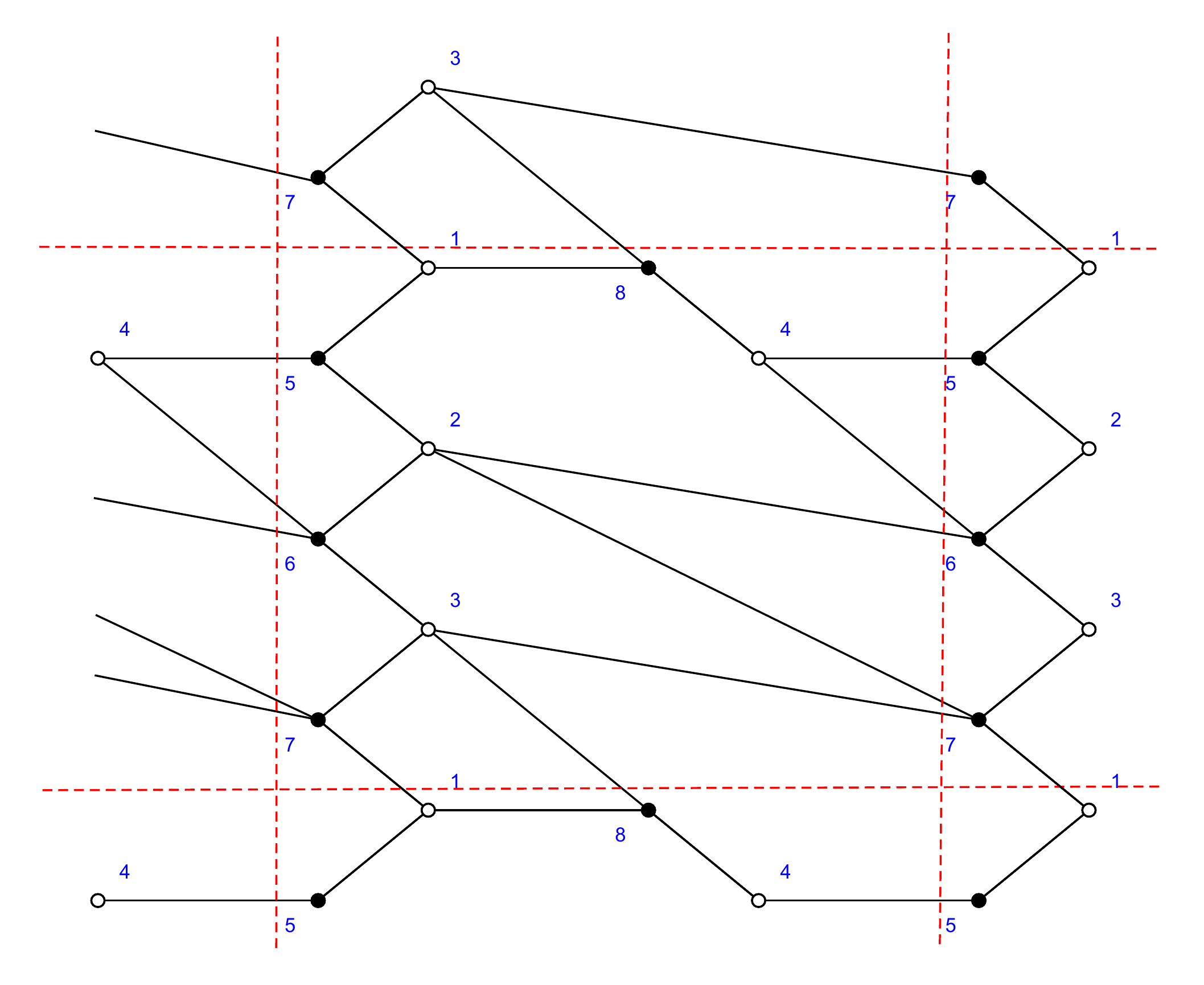}
\caption{$Y^{3,1}$}\label{fig:Y31}
\end{subfigure}
\begin{subfigure}{0.6\textwidth}
\includegraphics[width=\textwidth]{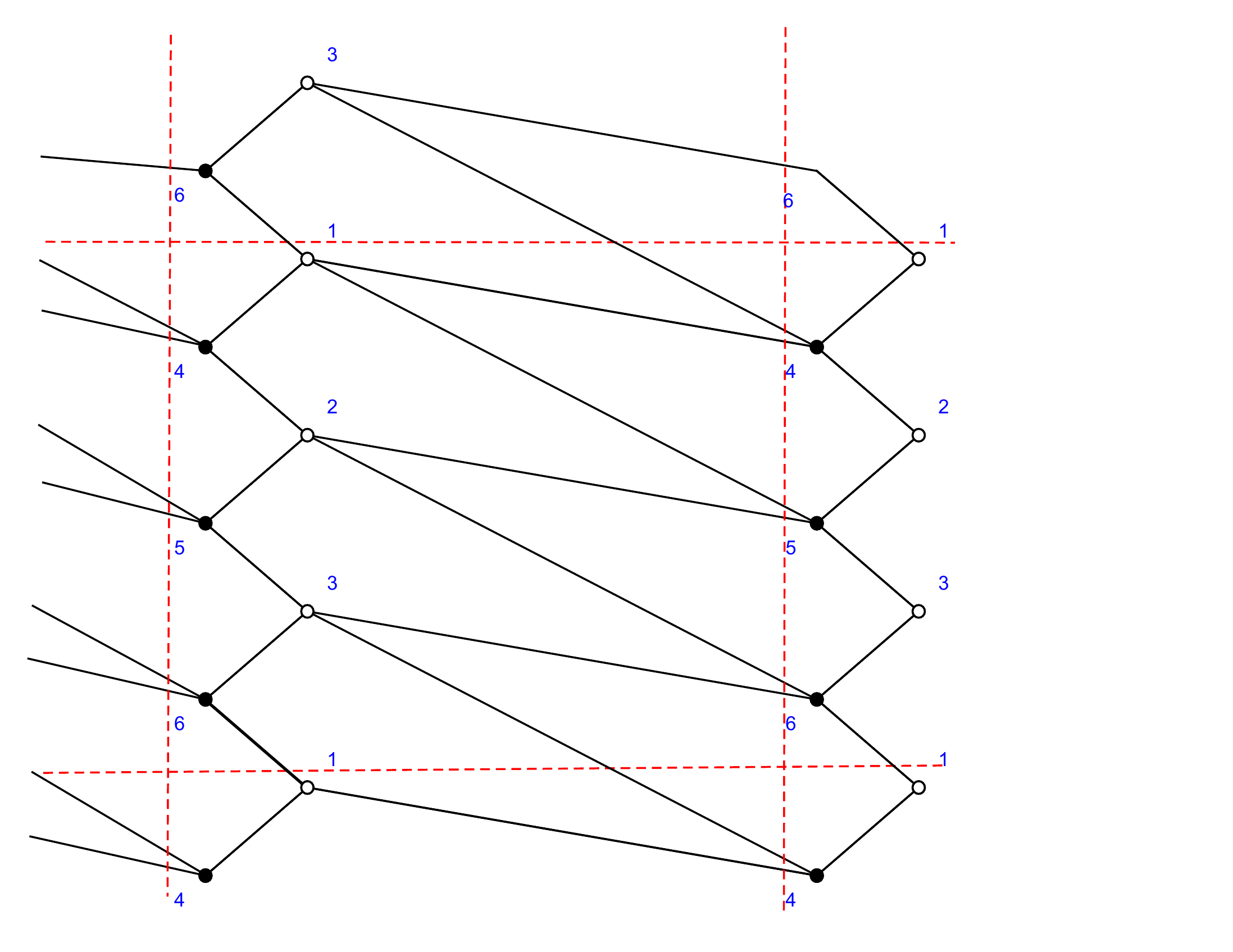}
\caption{$Y^{3,0}$}\label{fig:Y30}
\end{subfigure}
\caption{Brane tiling for $Y^{3,q},q=3,2,1,0$, the unit cells are divided by the red dashed lines, which are the loops $\gamma_{X,Y}$ on the torus}
\end{figure}

In this subsection, we give some examples for the dimer models of 5d $\mathcal{N}=1$ $SU(3)$ gauge theories with various Chern-Simons levels $m=0,1,2,3$. The graphs of these theories appear during the study of 4d $\mathcal{N}=1$ quiver gauge theories, where the graphs of the dimer models are brane tiling for the quiver gauge theories. For the $Y^{p,p}$ system, the brane tiling is the well-known Hexagon tiling \cite{Eager:2011dp}. In the quiver gauge theories, we can get $Y^{p,q},q<p$ theories by introducing impurities in the $Y^{p,p}$ quiver. Then we can get the brane tiling from the dual graph of the planer quiver. More technical details and examples can be found in \cite{Franco:2005rj, Benvenuti:2004dy,Benvenuti:2004wx}. The procedure in the quiver side can be alternatively understood by merging some points in the tiling for the $Y^{p,p}$ system to get the tiling for a $Y^{p,q},q<p$ system. For example, the tiling for $Y^{3,3}$ system is depicted in Figure \ref{fig:Y33}. One can get the brane tiling of $Y^{3,2}$ systems in Figure \ref{fig:Y32} by merging the point ${\color{blue}{8}},{\color{blue}{11}}$ and ${\color{blue}{2}},{\color{blue}{5}}$ in Figure \ref{fig:Y33}. By further merging ${\color{blue}{9}},{\color{blue}{12}}$ and ${\color{blue}{3}},{\color{blue}{6}}$, we get $Y^{3,1}$ \ref{fig:Y31}. By doing this further, we get $Y^{3,0}$ in Figure \ref{fig:Y30}. In the following, we list their Poisson brackets and Hamiltonians for these models. 

\subsection*{$Y^{3,3}$ model}
We choose the loops to be\footnote{There is an independent but irrelevant zig-zag path ${\color{blue}{1}}\rightarrow{\color{blue}{7}}\rightarrow{\color{blue}{2}}\rightarrow{\color{blue}{8}}\rightarrow{\color{blue}{3}}\rightarrow{\color{blue}{9}}\rightarrow{\color{blue}{1}}$ which commutes with other loops we choose. Since it is irrelevant for dynamical Hamiltonians, we don't mention it in other examples.} 
\be\label{Y33cycle}\begin{split}
&t_1={\color{blue}{7}}\rightarrow{\color{blue}{1}}\rightarrow{\color{blue}{10}}\rightarrow{\color{blue}{4}}\rightarrow{\color{blue}{7}}, 
\quad t_2={\color{blue}{7}}\rightarrow{\color{blue}{2}}\rightarrow{\color{blue}{11}}\rightarrow{\color{blue}{4}}\rightarrow{\color{blue}{7}}, \\
& t_3={\color{blue}{8}}\rightarrow{\color{blue}{2}}\rightarrow{\color{blue}{11}}\rightarrow{\color{blue}{5}}\rightarrow{\color{blue}{8}}, 
\quad t_4={\color{blue}{8}}\rightarrow{\color{blue}{3}}\rightarrow{\color{blue}{12}}\rightarrow{\color{blue}{5}}\rightarrow{\color{blue}{8}}, \\
& t_5={\color{blue}{9}}\rightarrow{\color{blue}{3}}\rightarrow{\color{blue}{12}}\rightarrow{\color{blue}{6}}\rightarrow{\color{blue}{9}}, 
\quad t_6={\color{blue}{9}}\rightarrow{\color{blue}{1}}\rightarrow{\color{blue}{10}}\rightarrow{\color{blue}{6}}\rightarrow{\color{blue}{9}}.\quad
\end{split}\ee
Only loops that are overlapped have non-vanishing Poisson brackets, they are
\be\label{Y33PB}\begin{split}
\{t_1,t_6\}=-t_1 t_6, &\quad \{t_1,t_2\}=t_1 t_2, \quad \{t_2,t_3\}=t_2 t_3,\\
\{t_3,t_4\}=t_3 t_4, &\quad \{t_4,t_5\}=t_4 t_5, \quad \{t_5,t_6\}=t_5 t_6.
\end{split}\ee
The Hamiltonians can be read from the graph directly from the rules in previous section, as the sum of one, two and three loops in the graph:
\be \label{Y33Hami}
\begin{split}
H_1&=t_1+t_2+t_3+t_4+t_5+t_6,\\
H_2&=t_1 t_3+t_1t_4+t_1 t_5+t_2t_4+t_2t_5+t_2t_6+t_3t_5+t_3t_6+t_4t_6,\\
H_3&=t_1t_3 t_5+t_2t_4t_6.
\end{split}\ee
Note that the number of independent Hamiltonians is equal to the genus of the mirror curves or the number of inner points in the toric diagram. Here $H_3$ is a Casimir instead of a Hamiltonian. It is important since there is a non-trivial instanton counting parameter, corresponding to the extra mass parameter among K\"ahler parameters.

The Poisson brackets (\ref{Y33PB}) can be enhanced to the quantum level as the commutation relations, in terms of canonical variables $\{q_i,p_i=-i\hbar \frac{\partial}{\partial q_i}\}$, we find a possible coordinates relation
\be\begin{split}
& t_1=R^2 e^{q_1},\quad t_2=e^{p_1+q_1},\quad t_3=R^2e^{q_2-q_1},\\ 
& t_4=e^{p_2+q_2},\quad t_5=R^2e^{-q_2},\quad t_6=e^{-p_1-p_2-q_1-q_2}. 
\end{split}\ee
The $R$ is the radius of the compactification circle from 5d to 4d, which gives a non-trivial deformation to the integrable systems. It is related to the instanton counting parameter or mass parameters in the 5d gauge theory point of view.

\subsection*{$Y^{3,2}$ model}
We choose the loops in Figure \ref{fig:Y32} 
\be\ba
&t_1={\color{blue}{6}}\rightarrow{\color{blue}{1}}\rightarrow{\color{blue}{9}}\rightarrow{\color{blue}{4}}\rightarrow{\color{blue}{6}},
\quad &t_2&={\color{blue}{6}}\rightarrow{\color{blue}{2}}\rightarrow{\color{blue}{7}}\rightarrow{\color{blue}{4}}\rightarrow{\color{blue}{6}}, \\
& t_3={\color{blue}{7}}\rightarrow{\color{blue}{2}}\rightarrow{\color{blue}{7}}, 
\quad &t_4&={\color{blue}{7}}\rightarrow{\color{blue}{3}}\rightarrow{\color{blue}{10}}\rightarrow{\color{blue}{2}}\rightarrow{\color{blue}{7}}, \\
 &t_5 ={\color{blue}{8}}\rightarrow{\color{blue}{3}}\rightarrow{\color{blue}{10}}\rightarrow{\color{blue}{5}}\rightarrow{\color{blue}{8}},
\quad &t_6&={\color{blue}{8}}\rightarrow{\color{blue}{1}}\rightarrow{\color{blue}{9}}\rightarrow{\color{blue}{5}}\rightarrow{\color{blue}{8}}.\quad
\ea\ee
The non-vanishing Poisson brackets are
\be\begin{split}
\{t_1,t_6\}=-t_1 t_6, &\quad \{t_1,t_2\}=t_1 t_2, \quad \{t_2,t_3\}=t_2 t_3, \quad \{t_2,t_4\}=t_2 t_4,\\
\{t_3,t_4\}=t_3 t_4, &\quad \{t_4,t_5\}=t_4 t_5, \quad \{t_5,t_6\}=t_5 t_6.
\end{split}\ee
In terms of canonical variables,
\be\ba
& t_1=R^2 e^{q_1},\quad t_2=e^{p_1+q_1},\quad t_3=R^2e^{q_2-q_1}, \\
& t_4=e^{p_2+q_2-q_1},\quad t_5=R^2e^{-q_2},\quad t_6=e^{-p_1-p_2-q_1}. 
\ea\ee

With the Hamiltonians
\be  \label{Y32Hami}
\begin{split}
H_1&=t_1+t_2+t_3+t_4+t_5+t_6,\\
H_2&=t_1 t_3+t_1t_4+t_1 t_5+t_2t_5+t_2t_6+t_3t_5+t_3t_6+t_4t_6,\\
H_3&=t_1t_3 t_5.
\end{split}\ee

\subsection*{$Y^{3,1}$ model}
We choose the loops in Figure \ref{fig:Y31} 
\be\ba
&t_1={\color{blue}{5}}\rightarrow{\color{blue}{1}}\rightarrow{\color{blue}{8}}\rightarrow{\color{blue}{4}}\rightarrow{\color{blue}{5}},
\quad &t_2&={\color{blue}{5}}\rightarrow{\color{blue}{2}}\rightarrow{\color{blue}{6}}\rightarrow{\color{blue}{4}}\rightarrow{\color{blue}{5}}, \\ 
& t_3={\color{blue}{6}}\rightarrow{\color{blue}{2}}\rightarrow{\color{blue}{6}}, 
\quad &t_4&={\color{blue}{6}}\rightarrow{\color{blue}{3}}\rightarrow{\color{blue}{7}}\rightarrow{\color{blue}{2}}\rightarrow{\color{blue}{6}}, \\
&t_5={\color{blue}{7}}\rightarrow{\color{blue}{3}}\rightarrow{\color{blue}{7}},
\quad &t_6&={\color{blue}{7}}\rightarrow{\color{blue}{1}}\rightarrow{\color{blue}{8}}\rightarrow{\color{blue}{3}}\rightarrow{\color{blue}{7}}.\quad
\ea\ee
The non-vanishing Poisson brackets are
\be\begin{split}
\{t_1,t_6\}=-t_1 t_6, &\quad \{t_1,t_2\}=t_1 t_2, \quad \{t_2,t_3\}=t_2 t_3, \quad \{t_2,t_4\}=t_2 t_4,\\
\{t_3,t_4\}=t_3 t_4, &\quad \{t_4,t_5\}=t_4 t_5, \quad \{t_4,t_6\}=t_4 t_6, \quad \{t_5,t_6\}=t_5 t_6.
\end{split}\ee
In terms of canonical variables, 
\be\ba
& t_1=R^2e^{q_1},\quad t_2=e^{p_1+q_1},\quad t_3=R^2e^{q_2-q_1}, \\
& t_4=e^{p_2+q_2-q_1},\quad t_5=R^2e^{-q_2},\quad t_6=e^{-p_1-p_2-q_1-q_2}.
\ea\ee
With the Hamiltonians
\be \label{Y31Hami}
\begin{split}
H_1&=t_1+t_2+t_3+t_4+t_5+t_6,\\
H_2&=t_1 t_3+t_1t_4+t_1 t_5+t_2t_5+t_2t_6+t_3t_5+t_3t_6,\\
H_3&=t_1t_3 t_5.
\end{split}\ee

\subsection*{$Y^{3,0}$ model}
We choose the loops in Figure \ref{fig:Y30} 
\be\ba
&t_1={\color{blue}{4}}\rightarrow{\color{blue}{1}}\rightarrow{\color{blue}{4}},
\quad &t_2&={\color{blue}{4}}\rightarrow{\color{blue}{2}}\rightarrow{\color{blue}{5}}\rightarrow{\color{blue}{1}}\rightarrow{\color{blue}{4}}, \\
&t_3={\color{blue}{5}}\rightarrow{\color{blue}{2}}\rightarrow{\color{blue}{5}}, 
\quad &t_4&={\color{blue}{5}}\rightarrow{\color{blue}{3}}\rightarrow{\color{blue}{6}}\rightarrow{\color{blue}{2}}\rightarrow{\color{blue}{5}},  \\
&t_5={\color{blue}{6}}\rightarrow{\color{blue}{3}}\rightarrow{\color{blue}{6}}, 
\quad &t_6&={\color{blue}{6}}\rightarrow{\color{blue}{1}}\rightarrow{\color{blue}{4}}\rightarrow{\color{blue}{3}}\rightarrow{\color{blue}{6}}.\quad
\ea\ee
The non-vanishing Poisson brackets are
\be\begin{split}
\{t_1,t_6\}=-t_1 t_6, &\quad \{t_1,t_2\}=t_1 t_2, \quad \{t_2,t_3\}=t_2 t_3,\quad \{t_3,t_4\}=t_3 t_4, \quad \{t_4,t_5\}=t_4 t_5,\\
 \quad \{t_5,t_6\}=t_5 t_6,& \quad \{t_2,t_4\}=t_2 t_4, \quad \{t_4,t_6\}=t_4 t_6, \quad \{t_2,t_6\}=-t_2 t_6. 
\end{split}\ee
In terms of canonical variables,
\be\ba
& t_1=R^2e^{q_1},\quad t_2=e^{p_1+q_1},\quad t_3=R^2e^{q_2-q_1},\\
& t_4=e^{p_2-q_1},\quad t_5=R^2e^{-q_2},\quad t_6=e^{-p_1-p_2}.
\ea\ee
With the Hamiltonians
\be
\begin{split} \label{Y30Hami}
H_1&=t_1+t_2+t_3+t_4+t_5+t_6,\\
H_2&=t_1 t_3+t_1t_4+t_1 t_5+t_2t_5+t_3t_5+t_3t_6,\\
H_3&=t_1t_3 t_5.
\end{split}\ee

\section{Perturbative computations of quantum spectra} \label{secperturbative} 
In this section, we consider the perturbative energy spectra of the quantum integrable systems described by genus two mirror curves, including the $Y^{3,m}$ models with $m=0,1,2,3$, and $\mathbb{C}^3/\mathbb{Z}_5$ model. Each model has two dynamical Hamiltonians, which are derived from dimer models. In the Section \ref{sec:dimermodel}, we derived the Hamiltonians for the $Y^{3,m}$ models, where the case of $m=0$ was also considered in \cite{HM1}. The Hamiltonians of some orbifold models including $\mathbb{C}^3/\mathbb{Z}_5$ are available in \cite{SHM1}. We also note that the $Y^{3,3}$ model is equivalent to the orbifold $\mathbb{C}^3/\mathbb{Z}_6$ model in \cite{SHM1}. We quantize the Hamiltonians by promoting the dynamical variables to operators with canonical commutation relations $[q_i, q_j]=[p_i, p_j]=0, [q_i, p_j]=i \hbar \delta_{i,j} $ with $i,j=1,2$.  

The Hamiltonians are bounded below in the phase space $(q_1,p_1, q_2, p_2)$. First we consider the $Y^{3,0}, Y^{3,3}, \mathbb{C}^3/\mathbb{Z}_5$ models, for which the classical minima are simply located at the origin $q_1=q_2=p_1=p_2=0$. We expand the Hamiltonians around the minimal point. 

First we study in details the $\mathbb{C}^3/\mathbb{Z}_5$ model, whose Hamiltonians are 
\begin{eqnarray} \label{HamiC3Z5}
H_1 &=& e^{q_1}+e^{p_1}+e^{-q_1 + q_2}+ e^{p_2}+e^{-q_2 - p_1 - p_2}, \\ 
H_2 &=& e^{q_2}+ e^{q_1 + p_2}+e^{p_1 + p_2}+ e^{ -p_2 - q_2}+e^{ -q_1 - p_1 - p_2}. 
\end{eqnarray} 
We expand the Hamiltonians up to quadratic order  
\begin{eqnarray} \label{secondorder2.1} 
H_i = 5 + \frac{1}{2} 
\begin{pmatrix} q_1 & q_2 & p_1 & p_2  
 \end{pmatrix} S_i \begin{pmatrix}
 q_1 \\ q_2 \\ p_1 \\ p_2  
 \end{pmatrix}  + \mathcal{O} (\hbar^{\frac{3}{2} } ) , ~~~~ i=1,2, 
 \end{eqnarray} 
 where the $S_1, S_2$ are real symmetric matrices 
 \begin{eqnarray} \label{Smatrix}
 S_1 = \begin{pmatrix} 2 & -1 & 0 & 0 \\
 -1 & 2 & 1 & 1\\
 0 & 1 & 2 & 1 \\
 0 & 1 & 1 & 2  
 \end{pmatrix} ,
 ~~ 
 S_2 = \begin{pmatrix} 2 & 0 & 1 & 2 \\
 0 & 2 & 0 & 1\\
 1 & 0 & 2 & 2 \\
 2 & 1 & 2 & 4  
 \end{pmatrix} . 
 \end{eqnarray}

We would like to write the quadratic Hamiltonians as linear combinations of two harmonic oscillators. We consider a linear transformation 
\begin{eqnarray}
 \begin{pmatrix}
 q_1 \\ q_2 \\ p_1 \\ p_2  
 \end{pmatrix} 
 = M 
 \begin{pmatrix}
 x_1 \\ x_2 \\ y_1 \\ y_2  
 \end{pmatrix} ,
\end{eqnarray} 
where $M$ is a $4\times 4$ real matrix. To preserve the same canonical commutation relation, the matrix $M$ must be a symplectic matrix $M\Sigma M^T =\Sigma$, where $\Sigma$ is the antisymmetric matrix 
\begin{eqnarray}
 \Sigma = \begin{pmatrix}
 0 & 0 & 1 & 0 \\ 0 & 0 & 0 & 1 \\ -1 & 0 & 0 & 0 \\ 0 & -1 & 0 & 0 
 \end{pmatrix} . 
\end{eqnarray} 
It turns out due to the special property that the Hamiltonians commute with each other, we can find symplectic transformation $M$ so that the quadratic terms can be written as linear combinations of the two harmonic oscillators 
\begin{eqnarray}
H_1 &=& 5+ \frac{1}{2} [c_1(x_1^2+y_1^2) + c_2(x_2^2 +y_2^2) ] + \mathcal{O} (\hbar^{\frac{3}{2} } ) , \nonumber \\
H_2 &=& 5+ \frac{1}{2} [c_3( x_1^2+y_1^2) + c_4 (x_2^2 +y_2^2) ] + \mathcal{O} (\hbar^{\frac{3}{2} } ). 
\end{eqnarray} 
There is a continuous 2-parameter family of solutions for the matrix $M$. Without loss of generality, we can use a particular solution 
\begin{eqnarray} 
{\footnotesize
M_{\mathbb{C}^3/\mathbb{Z}_5 } = \begin{pmatrix}
-\frac{(5-2 \sqrt{5})^{\frac{1}{4}} }{\sqrt{10}} & -\frac{(10-2\sqrt{5})^{\frac{1}{4}} }{\sqrt{5}} &
  (\frac{1}{4}+\frac{1}{2 \sqrt{5}})^{\frac{1}{4}} & 0 \\
 -\sqrt{\frac{2}{5}} (5-2 \sqrt{5})^{\frac{1}{4}} & -(\frac{13}{40}+\frac{29}{40 \sqrt{5}})^{\frac{1}{4}} & 0 &
  \frac{1}{2^{3/4}} (1+\frac{1}{\sqrt{5}})^{\frac{1}{4}}  \\
 -\frac{(10+2\sqrt{5} )^{\frac{1}{4}} }{\sqrt{5}} & (\frac{1}{20}+\frac{1}{10 \sqrt{5}})^{\frac{1}{4}} & 0 &
  -\frac{1}{\sqrt{2}} (1 - \frac{2 \sqrt{5}}{5} )^{\frac{1}{4}} \\
 (\frac{1}{8}+\frac{11}{40 \sqrt{5}})^{\frac{1}{4}} & -\frac{(25-11 \sqrt{5})^{\frac{1}{4}} }{2^{3/4} \sqrt{5}} &
  -\frac{1 }{2^{3/4}}  (1- \frac{\sqrt{5}}{5} )^{\frac{1}{4}} & -\frac{1 }{2^{3/4}}  (1+ \frac{\sqrt{5}}{5} )^{\frac{1}{4}} \\
   \end{pmatrix}, }
\end{eqnarray} 
with the linear coefficients 
\begin{eqnarray} \label{coefficients2.6}
&& c_1 = (\frac{5 + \sqrt{5}}{2})^{\frac{1}{2}} , ~~~~~ c_2 = (\frac{5 - \sqrt{5}}{2})^{\frac{1}{2}} , \nonumber \\
&& c_3 = (5- 2\sqrt{5})^{\frac{1}{2}} , ~~~~~ c_4 = (5 + 2\sqrt{5})^{\frac{1}{2}}  . 
\end{eqnarray} 

Denoting the quantum levels of the harmonic oscillators $(x_1, y_1)$ and $(x_2, y_2)$ by two non-negative integers $n_1, n_2$, the quantum spectrum up to order $\hbar$ is 
\begin{eqnarray} \label{C3Z5DIrect}
\begin{pmatrix}
 E_1 \\ E_2  
 \end{pmatrix}
 = 5 \begin{pmatrix}
 1 \\ 1  \end{pmatrix} +  
 \begin{pmatrix}
 c_1 & c_2 \\ 
 c_3 & c_4  
 \end{pmatrix} 
 \begin{pmatrix}
 n_1+\frac{1}{2} \\ n_2+\frac{1}{2} 
 \end{pmatrix} \hbar +\mathcal{O}(\hbar^2) .
\end{eqnarray} 

We can further compute the higher-order corrections to the energy spectra. We use the time-independent perturbation theory well-known in quantum mechanics, which separates a Hamiltonian into a zero-order part and a perturbation part 
\begin{eqnarray}
\mathcal{H} = \mathcal{H}_0 +\mathcal{H}^\prime, 
\end{eqnarray}
where the zero-order part $\mathcal{H}_0$ corresponds to the Hamiltonians up to quadratic order in (\ref{secondorder2.1}), while the perturbation part $\mathcal{H}^\prime$ corresponds to the higher-order terms. 

We denote the harmonic quantum states of the zero order Hamiltonians as $|n_1, n_2\ket$. Then the first few order corrections to energy spectra are
\begin{eqnarray} \label{perturbation2.9}
E_{(n_1,n_2)} &=& E^{(0)}_{(n_1,n_2)} + \bra n_1, n_2 | \mathcal{H}^{\prime} | n_1, n_2\ket \nonumber \\
&& + \sum_{(m_1,m_2)\neq (n_1, n_2)} \frac{ |\bra m_1, m_2 | \mathcal{H}^{\prime} | n_1, n_2\ket |^2 }{E^{(0)}_{(n_1,n_2)} - E^{(0)}_{(m_1,m_2)}}+\cdots .  
\end{eqnarray} 
To compute the next $\hbar^2$ order corrections, we need to expand the exponentials in the Hamiltonians (\ref{HamiC3Z5}) to cubic and quartic orders, and rewrite the canonical coordinates in terms of the standard creation and annihilation operators. For the first correction $\bra n_1, n_2 | \mathcal{H}^{\prime} | n_1, n_2\ket$, the cubic terms have no contribution since there is an odd number of creation and annihilation operators, while the quartic terms make an order $\hbar^2$ contribution. The cubic terms have a $\hbar^2$ order contribution in the more complicated second correction term in the above equation (\ref{perturbation2.9}). After some complicated calculations, we find the $\hbar^2$ order contributions to the quantum spectra. For the $\mathbb{C}^3/\mathbb{Z}_5$ model, the results are 
\begin{eqnarray} \label{spectrum2.10}
E_1 &=& 5 + [ (\frac{5 + \sqrt{5}}{2})^{\frac{1}{2}} n_1 + (\frac{5 - \sqrt{5}}{2})^{\frac{1}{2}} n_2+ \frac{1}{2} (5 + 2 \sqrt{5})^{\frac{1}{2}} ]\hbar \nonumber \\ &&  
+ [7 + 2 (3 + \sqrt{5}) n_1 (1 + n_1) +2 (3 - \sqrt{5}) n_2 (1 + n_2) ] \frac{\hbar^2}{40} +\mathcal{O}(\hbar^3),  \\ \nonumber 
E_2 &=& 5 + [ (5- 2\sqrt{5})^{\frac{1}{2}} n_1 + (5 + 2\sqrt{5})^{\frac{1}{2}}  n_2+ (\frac{5 + \sqrt{5}}{2} )^{\frac{1}{2}} ]\hbar  +  
 [ 3 + \sqrt{5}  + 4 n_1 \\ \nonumber && + 2 (2 - \sqrt{5}) n_1^2 + 4 (1 + \sqrt{5}) n_2 + 
  2 (2 + \sqrt{5}) n_2^2 + 4\sqrt{5} n_1n_2 ] \frac{\hbar^2}{20}  +\mathcal{O}(\hbar^3).
\end{eqnarray} 

It is well known that the eigenvalues of a matrix do not change under a similarity transformation of the matrix. Here analogously we find that the spectra in (\ref{spectrum2.10}) are independent of the choice of symplectic transformation, up to the trivial freedom of exchanging the two quantum numbers $n_1\leftrightarrow n_2$. This is easy to understand from the physics point of view since the Hamiltonians are the same regardless of the choices of the canonical coordinates. Furthermore, the linear coefficients (\ref{coefficients2.6}) are indeed related to the eigenvalues of certain matrices. We note that for a general even-dimensional real symmetric matrix $S$, since $\det(S\Sigma - \lambda I) = \det(\Sigma S- \lambda I)= \det(S\Sigma + \lambda I) = \det(\Sigma S + \lambda I) $, the eigenvalues of $S\Sigma$ and $\Sigma S$ are the same and always come in pairs with opposite signs. In our context, we find that for the matrices (\ref{Smatrix}) in the quadratic Hamiltonians, the eigenvalues of $S_1\Sigma$ and $S_2\Sigma$ are always purely imaginary and the positive imaginary parts are exactly the linear coefficients (\ref{coefficients2.6}). Namely, the eigenvalues of $S_1\Sigma$ are $\pm ic_1, \pm ic_2$ and the eigenvalues of $S_2\Sigma$ are $\pm ic_3, \pm ic_4$. This is also true for the $Y^{3,0}$ and $Y^{3,3}$ models discussed below. In Appendix \ref{AppendixEigenvalue} we give a simple general mathematical proof of this property.

Similarly we find the symplectic transformations and the perturbative energy spectra for the $Y^{3,0}$ and $Y^{3,3}$ models in (\ref{Y30Hami}, \ref{Y33Hami}). Again there is a continuous 2-parameter family of solutions for symplectic transformations. For the $Y^{3,0}$ model, we can use for example a solution 
\begin{eqnarray} {\scriptsize 
M_{Y^{3,0} } = \frac{1}{\sqrt{2R } 3^{\frac{1}{4}} (4 + R^2)^{\frac{1}{4}}} \begin{pmatrix}
-R & -2 R & -\sqrt{3} R & 0 \\
 -2 R & -R & 0 & -\sqrt{3} R \\
 \sqrt{R^2+4}-R & - \frac{\sqrt{R^2+4} + R }{2} & 0 & -\frac{ \sqrt{3}}{2}
  (\sqrt{R^2+4}+R) \\
 \frac{3 R-\sqrt{R^2+4} }{2} & \sqrt{R^2+4} & -\frac{\sqrt{3}}{2} (\sqrt{R^2+4}+R ) &
  \sqrt{3} R 
   \end{pmatrix} , } \nonumber 
\end{eqnarray} 
and the perturbative energy spectra are 
\begin{eqnarray} \label{spectrum4.13}
\begin{pmatrix}
 E_1 \\ E_2  
 \end{pmatrix} 
 &=& 3(1+R^2) \begin{pmatrix} 1 \\ 1  \end{pmatrix} 
 + \frac{\sqrt{3}R }{2} [ \sqrt{4+R^2} (  n_1 + n_2+ 1) \begin{pmatrix} 1 \\ 1  \end{pmatrix} + R (n_1-n_2) \begin{pmatrix} 1 \\ -1  \end{pmatrix}
  ]\hbar  \nonumber \\ &&
+ \big{\{} \big{[}4 (11+ 15 n_1 +6 n_1^2  +15 n_2 +6 n_2^2 +18 n_1n_2) +[ 5+6( n_1 +n_1^2+n_2 +n_2^2)] R^2
 \nonumber \\ && - \frac{72}{4+R^2} (2 +3n_1 +n_1^2  + 3n_2 +n_2^2 +4 n_1 n_2 ) \big{]} \begin{pmatrix} 1 \\ 1  \end{pmatrix}  \nonumber \\ && +6 R \sqrt{4+R^2}(n_1+ n_1^2 - n_2 - n_2^2 ) \begin{pmatrix} 1 \\ -1  \end{pmatrix} \big{\}} \frac{\hbar^2 }{72}  +\mathcal{O}(\hbar^3).
\end{eqnarray} 
We see there is an apparent symmetry of the spectra. The spectra of the two Hamiltonians $E_1\leftrightarrow E_2$ are exchanged if the quantum levels are exchanged $n_1\leftrightarrow n_2$. 

For the $Y^{3,3}$ model, the results are 
 \begin{eqnarray}
M_{Y^{3,3} } = 2 \cdot 3^{\frac{1}{4}} \sqrt{R} \begin{pmatrix}
-R & -2 R & -\sqrt{3} R & 0 \\
 -2 R & -R & 0 & -\sqrt{3} R \\
 R+2 & 2 R-1 & \sqrt{3} R & -\sqrt{3} \\
 2 R-1 & R+2 & -\sqrt{3} & \sqrt{3} R \\   \end{pmatrix} ,  \nonumber 
\end{eqnarray} 
\begin{eqnarray} \label{Y33Spectrum}
 E_1 &=& 3(1+R^2) + \sqrt{3}R (n_1+n_2+1) \hbar + 
 [ \frac{1 + R^2}{9} + \frac{n_1}{12} (1 + n_1)(1 + R + R^2) \nonumber \\ && + 
 \frac{n_2}{12} (1 + n_2) (1 - R + R^2) ] \hbar^2  +\mathcal{O}(\hbar^3),  \\ 
  E_2 &=& 3(1+R^2 +R^4) + \sqrt{3}R [ (n_1+n_2+1) (1+R^2) -(n_1-n_2) R] \hbar  \nonumber \\ && + 
  \big{\{} [4 + 3 (n_1 + n_1^2 + n_2 + n_2^2)](1 + R^4) - 
  9 (n_1 - n_2) (1 + n_1 + n_2) R (1 + R^2) \nonumber \\ &&  
  +  2 (8 + 15 n_1 + 6 n_1^2 + 15 n_2 + 6 n_2^2 + 18 n_1 n_2) R^2 \big{\}} \frac{\hbar^2}{36} 
  + \mathcal{O}(\hbar^3). \nonumber 
\end{eqnarray} 
There is also an apparent symmetry that under a T-duality like transformation $R\rightarrow \frac{1}{R}$, the energy spectra transforms as $E_1\rightarrow \frac{E_1}{R^2}, E_2\rightarrow \frac{E_2}{R^4}$. 

We need to be careful with a potential subtlety of perturbation theory here. For the first Hamiltonian of the $Y^{3,3}$ model, we see that the energy $E_1$ is degenerate up to $\hbar$ order for quantum states with the same $n_1+n_2$. It turns out that this does not affect the calculations in the formula (\ref{perturbation2.9}), as we check that the off-diagonal elements of the perturbation in the degenerate space actually vanish, i.e. $ \bra n_1+k, n_2-k | \mathcal{H}^{\prime} | n_1, n_2\ket =0 $ for $k=\pm 1, \pm 2$. The vanishing is trivial for cubic terms in the perturbation $ \mathcal{H}^{\prime}$, while we check by an explicit computation that it is also true for quartic terms. 

For the remaining $Y^{3,1}$ and $Y^{3,2}$ models (\ref{Y31Hami}, \ref{Y32Hami}), we need to determine the classical minima by solving for the critical points of Hamiltonians $\partial_{q_i} H = \partial_{p_i} H=0$ for $i=1,2$. We find that the minima are located at the same points for the two Hamiltonians of the quantum system due to the special property that the Hamiltonians commute with each other. In these models, it is much more complicated to find the symplectic transformations that diagonalize the quadratic terms of the Hamiltonians expanded around the minima. However, we can still use the formula in Appendix \ref{AppendixEigenvalue} to compute the $\hbar$-order contributions to the quantum spectra in terms of the eigenvalues of certain matrices from the quadratic terms.

For the $Y^{3,1}$ model, the minima are at 
\begin{eqnarray}
q_1 = -3\log(r), ~~~ q_2 = -6 \log(r), ~~~p_1 = -\log(r), ~~~p_2 = -\log(r + R^2), 
\end{eqnarray} 
where $r$ is the only positive root of the polynomial equation,
\be\ba
r^9+R^2 r^8=1,
\label{RandrY31}
\ea\ee
with numerical value, e.g., $r=0.921599$ for $R=1$. The quantum spectra are 
\begin{eqnarray}\label{Y31Spectrum}
E_1 &=& \frac{3 r + 4 R^2}{r^6 (r + R^2)} + 
r^2R [ (3r+2R^2 +2R \sqrt{r+R^2} )^{\frac{1}{2}} (n_1+\frac{1}{2}) \nonumber \\ && + (3r+2R^2 -2R \sqrt{r+R^2} )^{\frac{1}{2}} (n_2+\frac{1}{2})] \hbar+
 \mathcal{O}(\hbar^2),  \\ 
E_2 &=& \frac{3 r + 2 R^2}{r^4} + r^{\frac{9}{2}} R (r+R^2)^{\frac{1}{2}} [ (3r+4R^2 - 4R \sqrt{r+R^2} )^{\frac{1}{2}} (n_1+\frac{1}{2}) \nonumber \\ && + (3r+4R^2 + 4R \sqrt{r+R^2} )^{\frac{1}{2}} (n_2+\frac{1}{2})] \hbar
+ \mathcal{O}(\hbar^2 ). \nonumber 
\end{eqnarray} 

For the $Y^{3,2}$ model, the minima are at 
\begin{eqnarray}
&& q_1 = - 3\log [r (r + R^2)] , ~~~ q_2 = -\log[r^2 (r + R^2)] , \nonumber \\ &&
p_1 = -\log(r), ~~~ p_2 = -\log(r + R^2), 
\end{eqnarray} 
where $r$ is now the only positive root of the equation
\be\ba
r^9+4 R^2 r^8+6 R^4 r^7 +4 R^6 r^6 +R^8 r^5 =1,
\label{RandrY32}
\ea\ee
with numerical value, e.g., $r=0.665055 $ for $R=1$. The quantum spectra are  
\begin{eqnarray}\label{Y32Spectrum}
E_1 &=& \frac{3 r + 5 R^2}{ r^7 (r + R^2)^6} + 
\frac{r^3 R(r+R^2)^2}{\sqrt{2}} [ (6r+5R^2 +R \sqrt{4r+5 R^2} )^{\frac{1}{2}} (n_1+\frac{1}{2}) \nonumber \\ && + (6r+5R^2 -R \sqrt{4r+5 R^2} ) ^{\frac{1}{2}} (n_2+\frac{1}{2})] \hbar+ \mathcal{O}(\hbar^2),  
\\
 E_2 &=& \frac{3 r^2 + 7 r R^2 + 5 R^4}{ r^4 (r + R^2)^4} +
 \frac{r^2 R(r+R^2)}{\sqrt{2}} \{ [r^3 (r + R^2)^2 (6 r^3 + 25 r^2 R^2 + 30 r R^4 + 10 R^6) \nonumber \\ && - R (5 r^2 + 10 r R^2 + 4 R^4) \sqrt{r(4r+5 R^2)} ]^{\frac{1}{2}} (n_1+\frac{1}{2})  +  [r^3 (r + R^2)^2 (6 r^3 + 25 r^2 R^2 \nonumber \\ && + 30 r R^4 + 10 R^6)  +R (5 r^2 + 10 r R^2 + 4 R^4) \sqrt{r(4r+5 R^2)} ]^{\frac{1}{2}} (n_2+\frac{1}{2}) \}\hbar + \mathcal{O}(\hbar^2). \nonumber 
\end{eqnarray} 
Without solving the symplectic transformations for these two $Y^{3,1}$ and $Y^{3,2}$ models, there is an ambiguity of exchanging the quantum numbers $n_1\leftrightarrow n_2$ in the spectra. This can be fixed by comparing with the derivatives of periods of the corresponding Calabi-Yau geometries.

%%%%%%%%%%%%%%%%%%%%%%%%%%%%%%%%%%%%%%%%%%%%%%%%%

\section{From topological strings to energy spectra} \label{sectiontopo}
In this section, we will show that spectrum problems can be solved by utilizing the well-known methods in the topological string theory. 
More precisely, we calculate energy spectra by imposing the Bohr--Sommerfeld quantization conditions on the quantum B-periods of the mirror curves. 
First, we summarize some basic facts about classical/quantum mirror curves and general relations between the topological strings and the energy spectra.
After that, we will demonstrate how we calculate the energy spectra from the quantum periods in some concrete models.

\subsection{General aspects of classical/quantum curves}
We consider B model topological string theories on toric Calabi-Yau three-folds, where their topological information is captured by mirror curves.
A genus $g$ mirror curve is defined by an algebraic equation for $x,y\in \mathbb{C}$,
\be\ba
W(\re^x,\re^y;{\bm z})=0,
\label{curve}
\ea\ee
where ${\bm z}=(z_1,z_2,... ,z_s),~s\geq g$ are the complex structure moduli parameters. Generally, there are $g$ dynamical moduli corresponding to $g$ compact A- and B-cycles of the Riemann surface, and the $s-g$ remaining ones are known as non-dynamical mass parameters. We can define two kinds of classical periods called as A- and B-periods by integrating $y=y(x;{\bm z})$ around compact A-cycles and their dual B-cycles,
\be\ba
\Pi_{i}({\bm z}) = \oint_{A_i} y(x;{\bm z}) \rd x,
\qquad
\Pi_{i,d}({\bm z}) = \oint_{B_i} y(x;{\bm z}) \rd x ,
\qquad
i=1,...,g,
\label{ClaPeriod}
\ea\ee
where $y(x;{\bm z})$ is the solution of \eqref{curve}.

The mirror maps connecting the K\"ahler parameters with the complex structure moduli parameters can be written as linear combinations of the A-periods and the mass parameters
\be\ba
t_j({\bm z}) =\sum_{i=1}^g C_{ij} \Pi_i({\bm z}) + \textrm{mass terms} , ~~~~ j=1,2,\cdots s, 
\label{ClaMirror}
\ea\ee
where the mass terms depend only on logarithms of mass parameters and will not appear in quantum corrections. Here $C_{ij}$ is an intersection matrix of compact divisors and the base curves we have chosen. With a suitable choice of the base curves, parts of the $g\times s$ matrix $C_{ij}$ happen to be the Cartan matrix of the gauge group in the context of geometric realizations of gauge theories. %Parts of the $g\times s$ matrix $C$ will be identified with the Cartan matrix in the context of geometric realizations of gauge theories. 

The dual B-periods give the derivatives of the genus zero topological string amplitude with respect to the K\"ahler parameters, so-called prepotential $F_0({\bm t})$,
\be\ba \label{PrepotBperiod}
\Pi_{d,i}({\bm z}) =\frac{\del F_0({\bm t}({\bm z}))}{\del \Pi_{i}({\bm z})}=\sum_{j=1}^s C_{ij}\frac{\del F_0({\bm t}({\bm z}))}{\del t_j({\bm z})}, ~~~~ i=1,2,\cdots g,
\ea\ee
where ${\bm t} = (t_1,..., t_s)$.
From the prepotential, we define the Bohr-Sommerfeld volumes as the derivatives of prepotential with respect to $t_i$ with appropriate shift $4\pi^2b_i^{NS}$
\be\ba
\vol_i({\bm z}) =\sum_{j=1}^s C_{ij} \left(\frac{\del F_0({\bm t})}{\del t_j} +4\pi^2b_j^{NS}\right).
\label{ClaBSVol}
\ea\ee
This shift can be derived from the S-dual like invariance of the classical volumes \cite{Hatsuda:2015fxa}. It can be absorbed into the genus zero free energy by adding a $t_i$-linear term \cite{Sun:2016obh}. In gauge theory point of view, $b_i^{NS}$ comes from the one loop contribution. For 5d $\mathcal{N}=1$ pure $SU(N)$ gauge theories with Chern-Simons level, denoting the $t_{i},i\leq N-1$ the node of $A_N$ group, and $t_N$ the instanton counting parameter. By setting $b_N^{NS}=0$, we have $b_i^{NS}=b_{N-i}^{NS}=-\frac{(N-i)i}{12}$, for $i=1,2,\cdots, {\frac{N}{2}}$. For the $SU(3)$ models we consider, we always have $b_1^{NS}=b_2^{NS}=-\frac{1}{6},\, b_3^{NS}=0$.

For toric Calabi-Yau three-folds, an efficient way to calculate the A- and B-periods is to solve the Picard--Fuchs equations defined by
\be\ba
&\mathcal{L}_\alpha \Pi_{i}=0,\, \mathcal{L}_\alpha \Pi_{d,i}=0,
\\
&\cL_\alpha=\left[ \prod_{Q_{i}^\alpha>0} \left(\frac{\del}{\del x_i}\right)^{Q_{i}^\alpha} - \prod_{Q_{i}^\alpha>0} \left(\frac{\del}{\del x_i}\right)^{Q_{i}^\alpha}  \right],
\ea\ee
where $Q_i^{\alpha}$ are the charge vectors and $x_i$ are the homogeneous coordinates of the toric variety.
The differential operators $\cL_\alpha$ are known as the Picard--Fuchs operators.
The variables $x_i$ relate to ${\bm z}$ through the Batyrev coordinates,
\be\ba
z_\alpha = \prod_{i=1}^{k+3} x_i^{Q_i^\alpha}.
\ea\ee
The A- and B-periods correspond to logarithmic and double-logarithmic solutions.

Now we promote the classical variables $x, y$ to the quantum operators $\x, \y$ with the canonical commutation relation,
\be\ba
\left[\x, \y \right] = \ri \hbar.
\ea\ee
Accordingly, the mirror curve is replaced by the difference equation,
\be\ba
W(\re^\x, \re^\y)\Psi(x) =0,
\ea\ee 
where $\Psi(x)$ is a wave function of the quantum system.
We can solve the difference equation by utilizing the WKB analysis,
\be\ba
\Psi(x) = \exp \left( \frac{i}{\hbar} \int^x w(x^{\prime};\hbar) \rd x^{\prime} \right).
\ea\ee
Then, we can define quantum version of two periods, called as quantum A- and B-periods,
\begin{subequations}
\begin{align}
\Pi_{i}({\bm z}) &\to \Pi_{i}({\bm z};\hbar) = \sum_{n=0}^\infty \Pi_i^{(n)} \hbar^{n},  
\qquad
\Pi_i^{(n)} = \oint_A w^{(n)}(x) \rd x,
\\
\Pi_{d,i}({\bm z}) &\to \Pi_{i,d}({\bm z};\hbar) = \sum_{n=0}^\infty \Pi_{i,d}^{(n)} \hbar^{n},
\qquad
\Pi_{i,d}^{(n)} = \oint_B w^{(n)}(x) \rd x,
\label{QBperiod}
\end{align}
\end{subequations}
where we expand $w(x;\hbar)$ as a series in $\hbar$,
\be\ba
w(x;\hbar) = \sum_{n=0}^\infty w^{(n)} \hbar^n.
\ea\ee
In our example, $w^{(2n-1)},~n\in\mathbb{Z}_{>0}$ can be expressed as the total derivative of a simple function with no monodromy.
Thus, its contour integral vanishes, and only $\hbar^{2n}$-corrections survive.

The quantum corrected prepotential $F({\bm t};\hbar)$, so-called NS free energy, is defined by the NS limit of the refined topological string free energy,
\be\ba \label{FNS4.14}
F({\bm t},\hbar) = \sum_{n=0}^\infty F_n({\bm t}) \hbar^{2n}.
\ea\ee
Similar to the prepotential, the NS free energy satisfies following equation,
\be\ba
\Pi_{d,i}({\bm z};\hbar) =\sum_{j=1}^s C_{ij} \frac{\del F({\bm t}({\bm z};\hbar), \hbar)}{\del t_j({\bm z}, \hbar)} ,
\label{dualPeriod}
\ea\ee
where $t_{i}({\bm z};\hbar)$ are the quantum corrected mirror maps, so-called quantum mirror maps. %and $\Pi_{d,i}({\bm z};\hbar)$ are the quantum B-periods, which will be explained in a moment. 
Comparing both sides of \eqref{dualPeriod}, we can obtain the recursion relations which enable us to fix $F_i({\bm t})$ up to irrelevant constants and mass parameters.

The Bohr--Sommerfeld volumes \eqref{ClaBSVol} also have quantum corrections,
\be\ba
\vol_i({\bm z}) 
\to
\vol_i({\bm z};\hbar) =\sum_{n\geq0} \vol_i^{(2n)}({\bm z}) \hbar^{2n}.
\ea\ee
In quantum mechanics, the phase volume should be quantized.
In our case, the B-periods are quantized,
\be\ba
 \oint_{B_i} w(x,\hbar) \rd x= 2\pi \hbar \left(n_i+\frac{1}{2} \right).
 \qquad
 n_i \in \mathbb{Z}_{\geq0}.
\ea\ee
From \eqref{ClaBSVol} with the quantum corrections, we can rewrite the quantization conditions as follows,
\be\ba
\vol_i ({\bm z};\hbar) = 2\pi \hbar \left(n_i+\frac{1}{2} \right),
\quad
i=1,2,...,g.
\ea\ee
The dynamical complex structure moduli will correspond to Hamiltonians of quantum systems as we will see in concrete examples. As in the case of NS free energy, by expanding the quantum B-periods in $\hbar$, we can determine the quantum corrections to the energy eigenvalues recursively. The classical B-periods have to vanish at a classical minimal energy where this corresponds to the conifold point in the topological string moduli space.
Thus, to solve the spectral problems from the topological strings, we have to calculate the phase volumes at the conifold point.
It turns out that there is no logarithmic cut for the classical volumes (B-periods) at the conifold point, so they are the same as the quantum mirror maps up to numerical factors,
\be\ba
\vol_i(\textrm{Coni}; \hbar) \sim t_{i,c}(\textrm{Coni}; \hbar),
\label{volandmap}
\ea\ee 
where $\textrm{Coni}$ denotes the conifold point, and $t_{i,c}({\bm z}_{c};\hbar)$ are the quantum mirror maps expanded around the conifold point.
The numerical factors in the coefficients of $t_{i,c}({\bm z}_{c};\hbar)$ can be determined by comparing with the derivatives of the classical volumes at conifold point or the perturbative computations that we have done in the previous section.
Therefore, we can calculate the eigenvalues only by applying the quantum mirror maps near the conifold point.

Now we calculate the quantum periods.
It is straightforward to calculate the quantum A-periods from the definition by taking residues, whereas the direct computations of B-periods are usually not so easy.
Here we utilize the differential operator method proposed in \cite{MM}, and developed in \cite{Huang1}.

The important fact is that the quantum A-periods can be given by acting differential operators on the classical periods,
\be\ba\label{PeriodDiffOp}
\Pi_k ({\bm z}; \hbar) = \left(\sum_{n=0}^\infty \hbar^{2n} \cD_{2n} \right) \Pi_k({\bm z}), ~~ k=1,2,\cdots ,g,
\ea\ee
where 
\be\ba
\cD_{2n}=\cD_{2n} (\theta_{z_1},\theta_{z_2},...\theta_{z_s}),
\qquad
\theta_{z_i} = z_i \frac{\del}{\del z_i}.
\ea\ee
and the coefficients of $\theta_{z_i}$ are given by rational functions of $z_i$.
This means that we can obtain the differential operators in the conifold frame by transforming from large radius frame to the conifold frame, $z_i \to z_{c,i}$.
Then, by acting the differential operators on the classical A-periods expanded near the conifold point, we can obtain the quantum corrections in the conifold frame.
Since the mass parameters are annihilated by the differential operators, they do not receive the quantum corrections.

From (\ref{ClaMirror}) and \eqref{PeriodDiffOp}, the quantum mirror maps are determined by the same differential operators as
\be\ba
t_i ({\bm z}; \hbar) = \left(\sum_{n=0}^\infty \hbar^{2n} \cD_{2n} \right) t_i({\bm z}), ~~~ i =1,2,\cdots ,s. 
\label{QuantMirror}
\ea\ee
Interestingly, the differential operators that we treat in this paper do not depend on the choice of the cycles\footnote{It would be interesting to confirm this property in a general setup.}. 
Also, the classical mirror maps can be calculated from the Picard--Fuchs operators.
Therefore, it is enough to calculate one of the quantum A-periods to derive the differential operators and determine the quantum mirror maps.

By combining \eqref{volandmap} with \eqref{QuantMirror}, the quantum corrections to the volumes and their derivatives with respect to the eigenvalues are given by
\be\ba
&\partial^{p_1}_{E_1} \partial^{p_2}_{E_2}\cdots \partial^{p_s}_{E_s} \vol_{j}^{(2n)} \sim \partial^{p_1}_{E_1} \partial^{p_2}_{E_2}\cdots \partial^{p_s}_{E_s} (\cD_{2n} t_{c,j}),
\label{volandmapDiff}
\ea\ee
where $p_i\in \mathbb{Z}_{\geq0}$, $n\in\mathbb{Z}_{>0}$, and $j=1,2,...,s$.
To calculate the right hand side, we use $\partial_{E_i} = \sum_{j=1}^g (\partial_{E_i} z_{c,j}) \partial_{z_{c,j}}$.

Remarkably, this structure holds in the quantum B-periods; the quantum corrections to the B-periods can be calculated by acting above operators on the classical B-periods,
\be\ba
\Pi_{d,i} ({\bm z}; \hbar) = \left(\sum_{n=0}^\infty \hbar^{2n} \cD_{2n} \right) \Pi_{d,i}({\bm z}).
\ea\ee
This means that once we derive the differential operators $\cD_{2n}$ from the quantum A-periods that we know how to calculate systematically, we can obtain the quantum B-periods which are not easy to obtain by the direct computations of the cycle integrals.

Similar to previous paper \cite{Huang1}, we can derive recursion relations for the NS free energy by expanding the equations (\ref{FNS4.14}, \ref{dualPeriod}, \ref{QuantMirror}) in $\hbar$. We can explicitly do this for the first and second correction terms $F_{1,2}({\bm t})$, which are determined by the differential operators $\cD_2, \cD_4$. In our examples, the differential operators will be a linear combinations of first and second derivatives of the complex structure moduli. Suppose 
\begin{eqnarray} 
\cD_2 = \sum_i s^{(2)}_i \theta_i + \sum_{i,j} s^{(2)}_{i,j} \theta_i\theta_j, 
\end{eqnarray}  
where the coefficients $s_i, s_{i,j}$ are rational functions of complex structure moduli $z_i$'s. Denote the classical mirror maps as $t_i$, then it is straightforward to compute 
\begin{eqnarray}
\theta_k(\partial_{t_i}F_0) &=& \sum_j \theta_k(t_j) (\partial_{t_i} \partial_{t_j} F_0), \\ \nonumber 
\theta_k \theta_l (\partial_{t_i}F_0) &=& \sum_j \theta_k \theta_l (t_j) (\partial_{t_i} \partial_{t_j} F_0) + \sum_{j,m} \theta_k(t_j) \theta_l(t_m) (\partial_{t_i} \partial_{t_j}\partial_{t_m} F_0). 
\end{eqnarray} 
So we have 
\begin{eqnarray}
\cD_2 (\partial_{t_i}F_0) =\sum_j \cD_2(t_j) (\partial_{t_i} \partial_{t_j} F_0) + \sum_{j,k,l,m} s^{(2)}_{l,m} \theta_l(t_j) \theta_m(t_k) (\partial_{t_i} \partial_{t_j}\partial_{t_k} F_0). 
\end{eqnarray}
Combining the $\hbar^2$ equations of (\ref{FNS4.14}, \ref{dualPeriod}, \ref{QuantMirror}), we find the linear coefficients $s_i^{(2)}$ cancel out. The equation for first order NS free energy is then 
\begin{eqnarray}
\sum_{i=1}^sC_{ni}[ \partial_{t_i} F_1 - \sum_{j,k,l,m} s^{(2)}_{l,m} \theta_l(t_j) \theta_m(t_k) (\partial_{t_i} \partial_{t_j}\partial_{t_k} F_0) ]=0, ~~~ n=1,2,\cdots ,g.  
\label{F1andF0}
\end{eqnarray} 
If $s=g$ and the matrix $C_{ij}$ is invertible, it cancels out in the above equation. Otherwise, in general we need to solve the equations including the $C_{ij}$ matrix. 
Similarly, repeating the same computation to the next order, we have
\be\ba
& \sum_{i=1}^sC_{ni}[ \partial_{t_i} F_2 
- \sum_{j,k,l,m} s^{(4)}_{l,m} \theta_l(t_j) \theta_m(t_k) (\partial_{t_i} \partial_{t_j}\partial_{t_k} F_0)
-\sum_{j} \cD_2(t_j) (\partial_{t_i} \partial_{t_j} F_1)
\\
&\quad
-
\frac{1}{2} \sum_{j,k}
\cD_2(t_j) \cD_2(t_k)
( \partial_{t_i} \partial_{t_j} \partial_{t_k} F_0) ]=0, ~~~~~ n=1,2,\cdots ,g. 
\label{F2andF0}
\ea\ee
Again, the linear coefficients $s_i^{(4)}$ cancel out.
By using \eqref{F1andF0}, if the matrix $C_{ij}$ is invertible, we can eliminate $F_1$, and obtain the relation between $F_2$ and $F_0$.

\if0
In summary, we list the steps to obtain the energy eigenvalues, 
\begin{itemize}
\item[1.] We expand classical mirror maps around the conifold point where the Bohr--Sommerfeld volumes vanish.
\item[2.] We derive the differential operators from one of the quantum A-periods.
\item[3.] By expanding Bohr--Sommerfeld quantization conditions in a power series around $\hbar=0$, we obtain the recursion relation.
\item[4.] We obtain the quantum corrections to the Bohr--Sommerfeld volumes at conifold point by acting the differential operators derived in step 2 on the classical mirror map. 
The ambiguous numerical factors are fixed by comparing the usual perturbation theory.
\item[5.] By combining the result in step 3 and 4, we finally obtain the energy eigenvalues.
\end{itemize}
\fi

%%%%%%%%%%%%%%%%
\subsection{Examples}

In this section, we demonstrate the previous computations in some concrete models.
In our examples, we focus on the genus two mirror curves: $\mathbb{C}^3/\mathbb{Z}_5$ and $Y^{3,m}$ with $m=0,1,2,3$.
Most of the classical computations have already been done in, e.g., \cite{Sun:2016obh, CGM1, HM1, SHM1, KPMS}, and we gather the results to make the paper self-contained.
%In the first and second examples in the sections \ref{C3Z5} and \ref{Y30case}, we will use some notations which will be used in other models. 
In the following, we may omit some arguments in functions for short notation.

\subsubsection{$\mathbb{C}^3/\mathbb{Z}_5$ model}\label{C3Z5}

The mirror curve of $\mathbb{C}^3/\mathbb{Z}_5$ is defined by
\be\ba
&\re^x + \re^{-x+p} + \re^{-p} +z_1^{1/3} z_2 \re^{2x} + z_1^{-1/3} = 0.
\ea\ee
The Picard--Fuchs operators are
\be\ba
&\cL_1 = -2 \theta_1^2 \theta_2 + \theta_1^3 + z_1(-2 \theta_2 + 3 \theta_2^2 - \theta_2^3 + 6 \theta_1 -18 \theta_1 \theta_2 + 9 \theta_1 \theta_2^2 + 27 \theta_1^2 -27\theta_1^2 \theta_2 + 27 \theta_1^3),
\\
&\cL_2 = \theta_2^2 - 3 \theta_1 \theta_2 + z_2 (-2 \theta_2 -4 \theta_2^2 + \theta_1 + 4\theta_1 \theta_2 - \theta_1^2),
\\
&\cL_3 = \theta_1^2 \theta_2 + z_1 z_2 (-2 \theta_2^2 + 2\theta_2^3 + 7 \theta_1 \theta_2 -13 \theta_1 \theta_2^2 -3 \theta_1^2 +24 \theta_1^2 \theta_2 -9 \theta_1^3).
\label{PFeqC3Z5}
\ea\ee
To provide the solutions of the Picard--Fuchs equations, first we define following function,
\be\ba
\omega_0 (\rho_i) = \sum_{l,m\geq0} c(l,m;{\bm \rho}) z_1^{l+\rho_1}z_2^{m+\rho_2},
\ea\ee
where
\be
{\footnotesize \ba
c(l,m;{\bm \rho}) &= 
\frac{\Gamma(\rho_1+1)^2\Gamma(\rho_2+1)\Gamma(\rho_1 -2\rho_2 +1)\Gamma(-3 \rho_1 + \rho_2 +1)}{\Gamma(l+\rho_1+1)^2 \Gamma(m+\rho_2 +1)\Gamma(l-2m+\rho_1 -2\rho_2 +1)\Gamma(-3l + m -3\rho_1+\rho_2+1)}. 
\ea } \ee
We further define derivatives of $\omega_0 (\rho_i)$,
\be\ba
\omega_i = \frac{\partial \omega_0}{\partial \rho_i}\biggl|_{\rho_{1,2}=0},
\quad
\omega_{ij} = \frac{\partial^2 \omega_0}{\partial \rho_i \partial \rho_j}\biggl|_{\rho_{1,2}=0}.
\ea\ee
Then, the mirror maps are given by
\be\ba
t_1({\bm z}) = \omega_1 = \log z_1 -6 z_1 -z_2 +45 z_1^2 -\frac{3}{2} z_2^2 + \cO(z_i^3),
\\
t_2 ({\bm z}) = \omega_2 = \log z_2 +2 z_1 +2z_2 -15 z_1^2 +3 z_2^2 + \cO(z_i^3).
\ea\ee
The derivatives of the prepotential are
\be\ba
&\frac{\del F_0}{\del t_1} = 2\omega_{1,1} + 2\omega_{1,2} + 3\omega_{2,2},
\\
&\frac{\del F_0}{\del t_2} = \omega_{1,1} + 6\omega_{1,2} + 9\omega_{2,2}.
\ea\ee
The classical B-periods $\Pi_{d,i}~(i=1,2)$ are given by the formula (\ref{PrepotBperiod}), where the matrix $C_{ij}$ in this model is
\be\ba
C = 
\left[
\begin{array}{rr}
3 & -1 \\
-1 & 2
\end{array}
\right].
\label{CmatC3Z5}
\ea\ee

From the prepotential, the Bohr-Sommerfeld volumes are
\be\ba
&\vol^{(0)}_1({\bm z})=3\frac{\partial F_0}{\partial t_1} -\frac{\partial F_0}{\partial t_2}-\frac{\pi^2}{2} ,
\\
&\vol^{(0)}_2({\bm z})=-\frac{\partial F_0}{\partial t_2}+2\frac{\partial F_0}{\partial t_2} -\frac{2\pi^2}{3} ,
\label{BSC3Z5}
\ea\ee
where the complex structure moduli parameters $z_1, z_2$ are related to the eigenvalues of the quantum systems by 
\begin{eqnarray} \label{variables4.38}
z_1 = - \frac{E_1}{E_2^3} , ~~ z_2 =\frac{E_2}{E_1^2}. 
\end{eqnarray} 
The classical volumes should vanish at a conifold point, $z_1=-1/25,~z_2=1/5$, or $E_1=E_2=5$. We check this numerically.

Now let us consider the quantization of the mirror curve.
Accordingly, the classical mirror curve is replaced by following difference equation,
\be\ba
 \Psi(x+ \ri \hbar)+\re^{-x} \re^{\frac{i\hbar}{2}} \Psi(x- \ri \hbar) + \left( z_1^{\frac{1}{3}} z_2 \re^{2x} +\re^x + z_1^{-\frac{1}{3}} \right) \Psi(x)=0.
\ea\ee
According to \cite{ACDKV}, we can calculate the quantum A-periods by taking the residue,
\be\ba
\Pi({\bm z};\hbar) 
&= \frac{1}{5}\log(z_1^2z_2) +\oint_{x=-\infty} \rd x w(x;\hbar) 
\\
&= \frac{1}{5}\log(z_1^2z_2) - 6 z_1^2 z_2 +15 z_1^2 -2 z_1 - \left( 5 z_1^2 z_2-\frac{15 z_1^2}{2}+\frac{z_1}{4} \right) \hbar^2 + \cO(\hbar^4, z_i^3).
\ea\ee
We note that as familiar from literature, the logarithmic term is not captured by the residue calculations and is added by hand.

Let us express the coefficients $\Pi^{(n)}$ as the classical A-periods with the differential operators.
Since the differential operators giving $\Pi^{(n\geq4)}$ are tedious long expressions, here we provide the differential operator giving the leading correction to the classical periods as an example\footnote{We provide the results of differential operators giving higher order quantum corrections in the {\it mathematica} file. The results contain the differential operators of $\mathbb{C}^3/\mathbb{Z}_5$ and $Y^{3,m}$ with $m=0,1,2,3$. One can find it in the source file on the arXiv.},
\be\ba
 \cD_2 =\frac{1}{8} \theta_1^2 +\frac{1}{6} \theta_1 \theta_2 .
 \label{D2C3/Z5}
\ea\ee
By using the operator, we can obtain the leading correction to the quantum mirror maps $t^{(2)}_i({\bm z};\hbar)$ and the quantum B-periods $\Pi_{d,i}^{(2)}({\bm z};\hbar) $,
\be\ba
&t^{(2)}_i({\bm z};\hbar) =\cD_2 t_i({\bm z}),
\qquad
&\Pi_{d,i}^{(2)}({\bm z};\hbar) 
=\cD_2 \Pi_{d,i}^{(0)}({\bm z}),
\ea\ee
with $i=1,2$.

To check the consistency, we calculate the NS free energy near the large radius point. 
%By solving \eqref{dualPeriod} recursively, we find
%\be\ba
%&\del_{t_1}F_0 = \frac{1}{5} (2 \Pi_{d,1}^{(0)} +\Pi_{d,2}^{(0)}),
%\\
%&\del_{t_2}F_0 = \frac{1}{5} ( \Pi_{d,1}^{(0)}+3\Pi_{d,2}^{(0)}),
%\\
%&\del_{t_1}F_1 = \frac{1}{5} (2 \Pi_{d,1}^{(2)}+\Pi_{d,2}^{(2)} -5 (t_1^{(2)} \partial^2_{t_1} F_0+t_2^{(2)} \partial_{t_1} \partial_{t_2} F_0 )),
%\\
%&\del_{t_2}F_1 = \frac{1}{5} ( \Pi_{d,1}^{(2)}+3 \Pi_{d,2}^{(2)} -5 (t_2^{(2)} \partial^2_{t_2} F_0+t_1^{(2)} \partial_{t_1} \partial_{t_2} F_0 )),
%\\
%&\cdots.
%\ea\ee
By solving the recursion relations \eqref{F1andF0} and \eqref{F2andF0} with the matrix \eqref{CmatC3Z5} which is invertible, we find the NS free energy whose instanton parts $\left[ F_n\right]^{\text{inst.}}$ are given by
\be{\footnotesize \ba
\left[ F_1\right]^{\text{inst.}}
&=-\frac{127 Q_1^2 Q_2^2}{12}-\frac{1}{6} 65 Q_1^2 Q_2-\frac{129 Q_1^2}{16}+\frac{7 Q_1 Q_2^2}{8}+\frac{5 Q_1 Q_2}{6}+\frac{7 Q_1}{8}-\frac{Q_2^2}{12}-\frac{Q_2}{6}+ \cO(Q_i^3),
\\
\left[ F_2\right]^{\text{inst.}}
&=-\frac{2561 Q_1^2 Q_2^2}{720}-\frac{263 Q_1^2 Q_2}{72}-\frac{207 Q_1^2}{64}+\frac{29 Q_1 Q_2^2}{640}+\frac{67 Q_1 Q_2}{1440}+\frac{29 Q_1}{640}+\frac{Q_2^2}{180}+\frac{Q_2}{360}+ \cO(Q_i^3).
\ea } \ee
They agree with the topological vertex computations. 
%Of course, one can calculate $F_{1,2}$ by solving the recursion relations \eqref{F1andF0} and \eqref{F2andF0} since in this case the matrix $C_{ij}$ is square and invertible.

Now we are ready to calculate the quantum corrections to the energy spectra. 
The all-order Bohr--Sommerfeld quantization conditions in this case are given by
\be\ba
\vol_{i}(E_1, E_2;\hbar) = 2 \pi \hbar \left( n_i +\frac{1}{2} \right),\quad
i=1,2,
\label{AllOrderQCC3Z5}
\ea\ee
where $\vol_i(E_1, E_2;\hbar)$ are the quantum corrected phase volumes. 
To obtain the quantum corrected spectra, we define $E_i$ and $\vol_i(E_1, E_2;\hbar)$ as a series in $\hbar$,
\be\ba
&E_i = \sum_{n=0}^\infty E^{(n)}_{i} \hbar^n,
\\
&\vol_i(E_1, E_2;\hbar) = \sum_{n=0}^\infty \vol^{(2n)}_{i}(E_1,E_2) \hbar^{2n}.
\ea\ee
From the classical limit $\hbar =0$ of \eqref{AllOrderQCC3Z5}, the Bohr--Sommerfeld volumes vanish at the classical minimum energies,
\be\ba
E_1^{(0)}=5 =: E_{\m_1},
\qquad
E_2^{(0)}=5 =: E_{\m_2},
\ea\ee 
which corresponds to the conifold point. By expanding \eqref{AllOrderQC} in $\hbar$, we can obtain $E^{(n)}_{i}$ as functions of $\vol_i^{(n)}(E_{\m_1}, E_{\m_2})$, e.g.,
\be\ba \label{1stCorTop}
&E^{(1)}_{1}
=
\frac{2\pi \left\{ (n_1 +\frac{1}{2}) \partial_{E_2}\vol_2^{(0)} - (n_2 + \frac{1}{2}) \partial_{E_2}\vol_1^{(0)}\right\}}
{ \left\{ \partial_{E_1} \vol_1^{(0)} \partial_{E_2} \vol_2^{(0)} - \partial_{E_2} \vol_1^{(0)} \partial_{E_1} \vol_2^{(0)} \right\}},
\\
&E^{(1)}_{2}
=
\frac{2\pi \left\{ (n_1 +\frac{1}{2}) \partial_{E_1}\vol_2^{(0)} - (n_2 + \frac{1}{2}) \partial_{E_1}\vol_1^{(0)}\right\}}
{ \left\{ \partial_{E_1} \vol_1^{(0)} \partial_{E_2} \vol_2^{(0)} - \partial_{E_2} \vol_1^{(0)} \partial_{E_1} \vol_2^{(0)} \right\}},
\ea\ee
where we omit the arguments $(E_{\m_1}, E_{\m_2})$ of $\vol_i^{(n)}$ \footnote{We will use this expression for other models, where the arguments of $\vol_i^{(n)}$ in these models are $(E_{\m_1}, E_{\m_2},R)$.}.
By comparing \eqref{1stCorTop} with perturbative calculations (\ref{C3Z5DIrect}), we find the exact values of the $E_{1,2}$-derivatives of phase volumes at the classical minimum energies,
\be\ba
\begin{pmatrix}
\partial_{E_1} \vol_1^{(0)} & \partial_{E_2} \vol_1^{(0)}
\\
\partial_{E_1} \vol_2^{(0)} & \partial_{E_2} \vol_2^{(0)} 
 \end{pmatrix} 
 =
 \frac{\pi}{5}
 \begin{pmatrix}
 -2 \left( 5-2 \sqrt{5}\right)^{1/2} & \left( 2 \left(\sqrt{5}+5\right)\right)^{1/2} \\ 
 2 \left( 2 \sqrt{5}+5 \right)^{1/2} & - \left( 10-2 \sqrt{5} \right)^{1/2}
 \end{pmatrix}.
\ea\ee
With the change of variables \eqref{variables4.38}, we find
\begin{eqnarray}
\begin{pmatrix}
 \partial_{z_1}\vol^{(0)}_1({\bm z}) & \partial_{z_2}\vol^{(0)}_1({\bm z}) \\ 
  \partial_{z_1}\vol^{(0)}_2({\bm z}) & \partial_{z_2}\vol^{(0)}_2({\bm z})  
 \end{pmatrix} 
 =\pi
 \begin{pmatrix}
-10 \sqrt{5+2 \sqrt{5}} & \sqrt{130-58 \sqrt{5}} \\ 
-10 \sqrt{5-2 \sqrt{5}} & -\sqrt{130+58 \sqrt{5}} 
 \end{pmatrix} .
 \label{DzVolC3Z5}
\end{eqnarray}
We check this is indeed true numerically. 

The classical mirror maps near the conifold point can be obtained by solving the Picard--Fuchs equations \eqref{PFeqC3Z5} in the conifold frame,
\be{\footnotesize \ba
&t_{c,1} 
= -2 \pi \left( 5 \left(2 \sqrt{5}+5\right)\right)^{1/2} z_{c,1}- \pi \left( 26-\frac{38}{\sqrt{5}}\right)^{1/2}  z_{c,2} + 24 \pi \left( 1-\frac{2}{\sqrt{5}}\right)^{1/2} z_{c,1} z_{c,2}+ \cO(z_{c,i}^2),
\\
&t_{c,2}
=-2 \pi \left( 50-10 \sqrt{5}\right)^{1/2} z_{c,1}-2 \pi \left( 13+\frac{22}{\sqrt{5}}\right)^{1/2} z_{c,2} +24\pi \left( 2+\frac{2}{\sqrt{5}}\right)^{1/2} z_{c,1} z_{c,2} + \cO(z_{c,i}^2),
\label{mapC3Z5coni}
\ea } \ee
where
\be\ba
z_1 = \frac{1}{25} + z_{c,1},\quad
z_2 = - \frac{1}{5} + z_{c,2}.
\ea\ee
The coefficients of $z_{c,1}, z_{c,2}$ in the classical mirror map are fixed by the relation \eqref{volandmapDiff}.

We can calculate the next leading corrections to the energy spectra $E^{(2)}_{1,2}$ by looking at $\hbar^2$-order terms in \eqref{AllOrderQCC3Z5}.
To obtain them, we need to calculate the second derivatives of the volumes and leading corrections to the energy spectra.
The latter ones can be calculated from the formula \eqref{volandmapDiff} and explicit form of the differential operator \eqref{D2C3/Z5}.
After some computations, we find
\be{\footnotesize \ba
&E^{(2)}_{1} = \frac{1}{40} \left(-2 \left(\sqrt{5}-3\right) n_1 \left(n_1+1\right)+2 \left(\sqrt{5}+3\right) n_2 \left(n_2+1\right)+7\right),
\\
&E^{(2)}_{2} = \frac{1}{20} \left(2 \left(\sqrt{5}+2\right) n_1^2 -2 \left(\sqrt{5}-2\right) n_2^2 +4 \left(\sqrt{5}+1\right) n_1+4 n_2 + 4\sqrt{5} n_1 n_2 +\sqrt{5}+3\right).
\ea} \ee
These results agree with the perturbative computations \eqref{spectrum2.10}.

%%%%%%%%%%

\subsubsection{$Y^{3,0}$ model}\label{Y30case}

The mirror curve of $Y^{3,0}$ is given by
\be\ba
\re^{p}+z_1 z_2^2 z_3 \re^{-p+3x} + z_1 z_2^2 \re^{3x} +z_2 \re^{2x} + \re^{x}+1=0.
\ea\ee
The Picard--Fuchs operators are
\be\ba
&\cL_1=(\theta_1 - \theta_3)(\theta_1-2\theta_2) - z_1 (-2 \theta_1 + \theta_2 -1)(-2 \theta_1 + \theta_2),
\\
&\cL_2=(\theta_2 - \theta_3)(\theta_2 -2 \theta_1) - z_2 (-2 \theta_2 + \theta_1 -1)(-2 \theta_2 + \theta_1),
\\
&\cL_3 =\theta_3^2 - z_3 (\theta_1 - \theta_3)(\theta_2 - \theta_3),
\\
&\cL_4 = \theta_3^2 - z_1 z_2 z_3 (\theta_1- 2\theta_2)(\theta_2 - 2\theta_1).
\ea\ee 
Note that these operators are symmetric under exchange of $z_1$ and $z_2$.
To give the solutions of the Picard--Fuchs equations, we define following function,
\be\ba
\omega_0 (\rho_i) = \sum_{l,m,n\geq0} c(l,m,n;\rho_i) z_1^{l+\rho_1}z_2^{m+\rho_2} z_3^{n+\rho_3},
\label{omegaY30}
\ea\ee
where
\be\ba
c(l,m,n;\rho_i) &= 
\frac{1}{\Gamma(n+\rho_3+1)^2 \Gamma(-n+l+\rho_1-\rho_3+1)\Gamma(-n+m+\rho_2-\rho_3+1)}
\\ 
&\times
\frac{1}{\Gamma(l-2m+\rho_1-2\rho_2+1)\Gamma(-2l+m-2\rho_1+\rho_2+1)}.
\ea\ee
Then, the classical mirror maps and the derivatives of the prepotential are given by
\be\ba
&t_1({\bm z}) = \omega_1= \log z_1 + 2 z_1 + 3 z_1^2 - z_2 -\frac{3}{2} z_2^2 + z_1 z_2^2 - 2 z_1^2 z_2 -4 z_1^2 z_2 z_3 + 2 z_1 z_2^2 z_3 + \cO(z_i^3),
\\
&t_2({\bm z}) = \omega_2= t_1|_{z_1 \leftrightarrow z_2},
\\
&t_3({\bm z}) = \omega_3 =\log z_3,
\ea\ee
and
\be\ba
&\frac{\partial F_0}{\partial t_1} = \omega_{11} + \omega_{12} + \frac{1}{2} \omega_{22} + \frac{2}{3} \omega_{13} + \frac{1}{3} \omega_{23} + \frac{2\pi^2}{3},
\\
&\frac{\partial F_0}{\partial t_2} = \omega_{22} + \omega_{12} + \frac{1}{2} \omega_{11} + \frac{2}{3} \omega_{23} + \frac{1}{3} \omega_{13} + \frac{2\pi^2}{3},
\ea\ee
where
\be\ba
\omega_i = \frac{\partial \omega_0}{\partial \rho_i}\biggl|_{\rho_{1,2,3}=0},
\quad
\omega_{ij} = \frac{\partial^2 \omega_0}{\partial \rho_i \partial \rho_j}\biggl|_{\rho_{1,2,3}=0}.
\label{omega}
\ea\ee
The classical B-periods $\Pi_{d,i}~(i=1,2)$ are given by the formula (\ref{PrepotBperiod}) with the matrix
\be\ba\label{MatrixY30}
C = 
\left[
\begin{array}{rrr}
2 & -1 & 0 \\
-1 & 2 & 0
\end{array}
\right],
\ea\ee
where the first $2\times 2$ block is the Cartan matrix of SU(3). 
From the prepotential, the Bohr-Sommerfeld volumes are given by
\begin{eqnarray}
\vol^{(0)}_i({\bm z})=\sum_{j=1}^3C_{ij}\frac{\partial F_0}{\partial t_j} -\frac{2\pi^2}{3} ,~~~ i=1,2, 
\label{Y30BSVol}
\end{eqnarray}  
where the complex structure moduli parameters $z_1, z_2, z_3$ are related to the quantum systems by 
\begin{eqnarray}
z_1 =\frac{E_2}{E_1^2} , ~~ z_2 =\frac{E_1}{E_2^2}, ~~ z_3 =-R^6. 
\end{eqnarray} 
The Bohr-Sommerfeld volumes should vanish at the conifold point, $z_1=z_2= \frac{1}{3(1+R^2)}$. We check numerically this is indeed true for, e.g., $R=1$.  

Now let us consider the quantum mirror curve defined by
\be\ba
\Psi(x- \ri \hbar) +z_1 z_2^2 z_3 \re^{3x} \re^{\frac{3i\hbar}{2}} \Psi(x+ \ri \hbar) + \left( z_1 z_2^2 \re^{3x} + z2 \re^{2x} + \re^x + 1\right) \Psi(x) =0.
\ea\ee
By taking the residue of $ w(x;\hbar)$, we find a quantum A-period,
\be\ba
\Pi({\bm z};\hbar) 
&= -\frac{1}{3} \log(z_1 z_2^2)+ \oint_{x=\infty} \rd x w(x;\hbar) 
\\
&= -\frac{1}{3} \log(z_1 z_2^2) + \left(-z_2 -\frac{3 z_2^2}{2} -\frac{10 z_2^3}{3} +z_1 z_2^2 +4 z_1 z_2^3 +2 z_1 z_2^2 z_3 + 12 z_1 z_2^3 z_3 \right)
\\
&\qquad -\left(\frac{1}{4} z_1 z_2^2 z_3 + \frac{7}{2} z_1 z_2^3 z_3\right) \hbar^2
+ \cO(\hbar^4, z_i^4).
\ea\ee
The differential operator giving the first quantum correction is
\be\ba
 \cD_2
 = \frac{1}{12} z_1 z_2 (5 z_3+4) \theta_1^2 +\frac{1}{12} z_1 z_2 (5 z_3+4) \theta_2^2 + \frac{1}{24}(-20z_1 z_2 -25 z_1 z_2 z_3+4) \theta_1 \theta_2 .
\ea\ee
Then, we can obtain the $\hbar^2$-correction to the quantum mirror maps and the quantum B-periods by acting above differential operator on the classical periods,
\be\ba
t^{(2)}_i = \cD_2 t_i,
\qquad
{\Pi}_{d,i}^{(2)}
=\cD_2 \Pi_{d,i}^{(0)},
\qquad
i=1,2.
\ea\ee
We note that in this model, the $t_3$ depends only on mass parameter $R$ and does not receive quantum corrections.

To check the consistency, we calculate the NS free energy near the large radius point. 
As in the $\mathbb{C}^3/\mathbb{Z}_5$ model, we can obtain the NS free energy by solving the recursion relations \eqref{F1andF0} and \eqref{F2andF0} whose instanton parts $\left[ F_n\right]^{\text{inst.}}$ are given by
%By solving \eqref{dualPeriod} with \eqref{MatrixY30} recursively, we find
%\be\ba
%&\partial_{t_1}F_0 = \frac{1}{3}(2\Pi_{d,1}^{(0)}+\Pi_{d,2}^{(0)}),
%\\
%&\partial_{t_2}F_0 = \frac{1}{3}(\Pi_{d,1}^{(0)}+2\Pi_{d,2}^{(0)}),
%\\
%&\partial_{t_1}F_1 = \frac{1}{3}(2\Pi_{d,1}^{(2)}+\Pi_{d,2}^{(2)}-3(t_1^{(2)}\partial^2_{t_1}F_0 + t_2^{(2)} \partial_{t_1}\partial_{t_2} F_0))
%\\
%&\partial_{t_2}F_1 = \frac{1}{3}(\Pi_{d,1}^{(2)}+2\Pi_{d,2}^{(2)}-3(t_2^{(2)}\partial^2_{t_2}F_0 + t_1^{(2)} \partial_{t_1}\partial_{t_2} F_0))
%\label{NSY30}
%\ea\ee
% From them, we find the NS free energy whose instanton parts $\left[ F_n\right]^{\text{inst.}}$ are given by
\be {\footnotesize \ba
\left[ F_1\right]^{\text{inst.}}
&=-\frac{1}{6}Q_1 -\frac{1}{6}Q_2 -\frac{1}{12}Q_1^2 -\frac{1}{12}Q_2^2 - \frac{1}{6}Q_1 Q_2 - \frac{1}{24} Q_1 Q_2 Q_3
\\
&\quad
-\frac{1}{12}Q_1^2 Q_2^2+\frac{7}{8} Q_1^2 Q_2 Q_3+\frac{7}{8} Q_1 Q_2^2 Q_3 +\frac{5}{6} Q_1^2 Q_2^2 Q_3 - \frac{1}{48} Q_1^2 Q_2^2 Q_3^2 + \cO(Q_i^3),
\\
\left[ F_2\right]^{\text{inst.}}
&=\frac{1}{360}Q_1 + \frac{1}{360}Q_2 + \frac{1}{180}Q_1^2 + \frac{1}{180}Q_2^2 +\frac{1}{360}Q_1 Q_2 +\frac{7}{5760} Q_1 Q_2 Q_3
\\
&\quad
+\frac{1}{180}Q_1^2 Q_2^2 +\frac{29}{640}Q_1^2 Q_2 Q_3+\frac{29}{640}Q_1 Q_2^2 Q_3 +\frac{37}{1440}Q_1^2 Q_2^2 Q_3 +\frac{7}{2880}Q_1^2 Q_2^2 Q_3^2 + \cO(Q_i^3).
\ea } \ee
They agree with the topological vertex computations.
%Since in this model the first $2\times 2$ block of the matrix $C_{ij}$ is invertible, one can use the recursion relations \eqref{F1andF0} and \eqref{F2andF0} to obtain above results.
Accidentally, it turns out that the derivatives with respect to the mass parameter $\partial_{t_3}F_1$ also satisfy a similar equation although it does not formally appear in \eqref{F1andF0} for this model. 

Now we are ready to calculate the quantum corrections to the energy spectra. The all-order Bohr--Sommerfeld quantization conditions are given by
\be\ba
 \vol_{i}(E_1, E_2,R;\hbar) = 2 \pi \hbar \left( n_i +\frac{1}{2} \right),\quad
i=1,2,
\label{AllOrderQC}
\ea\ee
where $\vol_i(E_1, E_2,R;\hbar)$ are the quantum corrected phase volumes. 
To obtain the quantum corrected spectra, we define $E_i$ and $\vol_i(E_1, E_2,R;\hbar)$ as a series in $\hbar$,
\be\ba
&E_i = \sum_{n=0}^\infty E^{(n)}_{i} \hbar^n,
\\
&\vol_i(E_1, E_2,R;\hbar) = \sum_{n=0}^\infty \vol^{(n)}_{i}(E_1,E_2,R) \hbar^{2n}.
\label{expandEandVol}
\ea\ee
From the classical limit $\hbar =0$ of \eqref{AllOrderQC}, the Bohr--Sommerfeld volumes vanish at the classical minimum energies,
\be\ba
E_i^{(0)}= 3(1+R^2) =: E_\m,
\ea\ee 
which correspond to the conifold point.
In the following, we demonstrate the computations for $R=1$. 
The leading corrections to the spectra are given by \eqref{1stCorTop}.
By comparing them with direct perturbative calculations (\ref{spectrum4.13}), we obtain the exact values of $E_{1,2}$-derivatives of the volumes,
\be\ba
\begin{pmatrix}
\partial_{E_1} \vol_{1}^{(0)} & \partial_{E_2} \vol_{1}^{(0)}
\\
\partial_{E_1} \vol_{2}^{(0)} & \partial_{E_2} \vol_{2}^{(0)}
 \end{pmatrix} 
 =
 \pi
\begin{pmatrix}
\left( \frac{2}{15} \left(\sqrt{5}+3\right)\right)^{1/2} & - \left( \frac{2}{15} \left(3-\sqrt{5}\right)\right)^{1/2}
\\
 \left( \frac{2}{15} \left(3-\sqrt{5}\right)\right)^{1/2} & \left( \frac{2}{15} \left(3+\sqrt{5}\right)\right)^{1/2}
\end{pmatrix} .
\ea\ee
With the changes of variables, we find
\begin{eqnarray}
\begin{pmatrix}
 \partial_{z_1}\vol^{(0)}_1({\bm z}) & \partial_{z_2}\vol^{(0)}_1({\bm z}) \\ 
  \partial_{z_1}\vol^{(0)}_2({\bm z}) & \partial_{z_2}\vol^{(0)}_2({\bm z})  
 \end{pmatrix} 
 = -\frac{8\sqrt{3}\pi}{3-\sqrt{5}}
 \begin{pmatrix}
 1-\sqrt{5} & -5+2\sqrt{5} \\ 
 -5+2\sqrt{5} & 1-\sqrt{5} 
 \end{pmatrix},
\end{eqnarray}
They agree with the direct computations numerically.

To obtain the derivatives of the volumes, we use the classical periods near the conifold point,
\be {\footnotesize \ba
t_{c,1} 
=& -\frac{4\pi (5+9\sqrt{5})}{5\sqrt{3}} z_{c,1} + \frac{4\pi (5-9\sqrt{5})}{5\sqrt{3}} z_{c,2} + z_{c,3} + \frac{1312 \pi }{25\sqrt{15} } z_{c,1} z_{c,2} + \frac{8\pi (125+117\sqrt{5}) }{1125 \sqrt{3} } z_{c,1} z_{c,3} 
\\ &
- \frac{8\pi (125-117\sqrt{5}) }{1125 \sqrt{3} } z_{c,2} z_{c,3} - \frac{4544 \pi}{1125 \sqrt{15}} z_{c,1}z_{c,2}z_{c,3} + \cO(z_{c,i}^2),
\\
t_{c,2} =& t_{c,1}|_{z_{c,1}\leftrightarrow z_{c,2}},
\\
t_{c,3} =& \log(-1+z_{c,3}),
\label{mapY30coni}
\ea } \ee
where
\be\ba
z_1 = \frac{1}{6} + z_{c,1},\quad
z_2 = \frac{1}{6} + z_{c,2},\quad
z_3 = -1 + z_{c,3}.
\ea\ee
The coefficients of $z_{c,1}$ and $z_{c,2}$ are fixed by the relation \eqref{volandmapDiff}.

From them, we can obtain the next leading order of the quantum corrections to the energy spectra by looking at $\hbar^2$-order of \eqref{AllOrderQC}. 
After some computations, we find
\be {\footnotesize \ba
E^{(2)}_{1} =& \frac{1}{360} \biggl(6 (19+ 5 \sqrt{5} )n_1 +6 (19 -5 \sqrt{5} ) n_2
+ 6 (13+5 \sqrt{5} ) n_1^2 + 6 (13-5 \sqrt{5} ) n_1^2 + 72 n_1 n_2 +101\biggr),
\\
E^{(2)}_{2} =& E^{(2)}_{1}|_{n_1 \leftrightarrow n_2}.
\\
\ea } \ee
They agree with the perturbative computations \eqref{spectrum4.13}.

%%%%%%%%%%

\subsubsection{$Y^{3,1}$ model}
In this example, we sometimes use some of the notations and definitions in Section \ref{Y30case}.
The mirror curve of $Y^{3,1}$ is given by
\be\ba
\re^p + z_3 \re^{2x-p}+ z_1 z_2^2 \re^{3x} + z_2 \re^{2x}+\re^x +1 =0.
\ea\ee
The Picard--Fuchs operators are
\be\ba
&\cL_1=\theta_1 (\theta_1 -2 \theta_2 -2 \theta_3) -z_1 (-2\theta_1 + \theta_2 -1)(-2 \theta_1 + \theta_2),
\\
&\cL_2=\theta_2 (-2\theta_1 + \theta_2) - z_2 (\theta_1 -2 \theta_2 -2\theta_3 -1)(\theta_1 -2 \theta_2 -2\theta_3),
\\
&\cL_3 =\theta_3^2 - z_3 (\theta_1 - 2\theta_2 -2\theta_3 -1)(\theta_1 -2\theta_2 -2\theta_3),
\\
&\cL_4 =\theta_1 \theta_2 \theta_3^2 -z_1 z_2 z_3(-2 \theta_1 + \theta_2)(\theta_1 -2 \theta_2 -2\theta_3-2)(\theta_1 -2 \theta_2 -2\theta_3 -1 )(\theta_1 -2 \theta_2 - 2\theta_3) .
\ea\ee 
Then, the classical mirror maps and the derivatives of the prepotential are given by
\be\ba
t_1({\bm z}) = \omega_1=\log z_1 
&+2 z_1 -z_2 -z_3 +3 z_1^2 -\frac{3 z_2^2}{2} -\frac{3 z_3^2}{2} -6 z_2 z_3 -2 z_1^2 z_2+6 z_1^2 z_3 +z_1 z_2^2
\\
&
-30 z_2^2 z_3-30 z_2 z_3^2 -315 z_2^2 z_3^2 +12 z_1 z_2^2 z_3 +90 z_1 z_2^2 z_3^2 + \cO(z_i^3),
\\
t_2({\bm z}) = \omega_2=\log z_2
&-z_1 +2 z_2 +2 z_3 -\frac{3 z_1^2}{2} +3 z_2^2 +3 z_3^2 +12 z_2 z_3  +z_1^2 z_2-3 z_1^2 z_3-2 z_1 z_2^2
\\
&+60 z_2^2 z_3 +60 z_2 z_3^2 +630 z_2^2 z_3^2 -24 z_1 z_2^2 z_3 -180 z_1 z_2^2 z_3^2 + \cO(z_i^3),
\\
t_3({\bm z}) = \omega_3 =\log z_3
&+\frac{1}{3}\left\{ 2\left(t_1({\bm z}) -\log z_1 \right) +4 \left(t_2({\bm z}) -\log z_2 \right) \right\},
\ea\ee
and
\be\ba
&\frac{\partial F_0}{\partial t_1} =\frac{1}{9}\left(4 \omega_{11} -2\omega_{12} -2\omega_{22} + 6 \omega_{13} +3 \omega_{23} \right),
\\
&\frac{\partial F_0}{\partial t_2} =\frac{1}{9}\left(- \omega_{11} -4\omega_{12} -4\omega_{22} + 3 \omega_{13} +6 \omega_{23} \right),
\\
&\frac{\partial F_0}{\partial t_3} = \frac{1}{3} \left(\omega_{11} + \omega_{22} + \omega_{12}  \right),
\ea\ee
where $\omega_i$ and $\omega_{ij}$ are defined in \eqref{omega}, and $\omega_0$ are defined in \eqref{omegaY30} with the coefficients $c(l,m,n;\rho_i)$
\be\ba
c(l,m,n;\rho_i) = 
&\frac{\Gamma \left(\rho _1+1\right) \Gamma \left(\rho _2+1\right) \Gamma \left(\rho _3+1\right)^2}{\Gamma \left(l+\rho _1+1\right) \Gamma \left(m+\rho _2+1\right) \Gamma \left(n+\rho _3+1\right)^2 }
\\
&\times
\frac{\Gamma \left(-2 \rho _1+\rho _2+1\right) \Gamma \left(\rho _1-2 \rho _2-2 \rho _3+1\right)}{\Gamma \left(l+\rho _1-2 \left(m+\rho _2\right)-2 \left(n+\rho _3\right)+1\right)\Gamma \left(m-2 \left(l+\rho _1\right)+\rho _2+1\right)}.
\ea\ee
The classical B-periods $\Pi_{d,i}~(i=1,2)$ are given by the formula (\ref{PrepotBperiod}) with the non-invertible matrix,
\be\ba\label{MatrixY31}
C = 
\left[
\begin{array}{rrr}
2 & -1 & 0 \\
-1 & 2 & 2
\end{array}
\right].
\ea\ee
From the prepotential, the Bohr-Sommerfeld volumes are
\begin{eqnarray}
\vol^{(0)}_i({\bm z})=\sum_{j=1}^3C_{ij}\frac{\partial F_0}{\partial t_j} -\frac{2\pi^2}{3},~~~ i=1,2.
\label{Y31BSVol}
\end{eqnarray}  
The complex structure moduli $z_1, z_2, z_3$ are related to the eigenvalues of the dimer system by,
\begin{eqnarray}
z_1 =\frac{E_2}{E_1^2} , ~~ z_2 =\frac{E_1}{E_2^2}, ~~ z_3 =\frac{R^6}{E_2^2}. 
\end{eqnarray} 
From the classical limit $\hbar=0$ of \eqref{AllOrderQC}, the classical phase volumes vanish at a conifold point
\be\ba
z_1 = \frac{r^9+2}{(r^9-4)^2},~~ z_2 = -\frac{r^{18}(r^4-4)}{(r^9+2)^2},~~ z_3 = -\frac{(r^9-1)^3}{(r^9+2)^2},
\ea\ee
where we use the polynomial relation \eqref{RandrY31} to eliminate $R$.
As a consistency check, we check numerically that the phase volumes \eqref{Y31BSVol} vanish at the conifold point for $r=2^{-1/9}$.

Now let us consider the quantum mirror curve given by
\be\ba
\Psi(x- \ri \hbar) + \re^{\ri \hbar} z_3 \re^{2x} \Psi(x+ \ri \hbar) + \left( z_1 z_2^2 \re^{3x} + z_2 \re^{2x}+\re^x +1\right) \Psi(x) =0.
\ea\ee
By taking the residue of $ w(x;\hbar)$, we find a quantum A-period,
\be\ba
\Pi({\bm z};\hbar) 
&= -\frac{1}{3} \log(z_1 z_2^2) +\oint_{x=\infty} \rd x \, w(x;\hbar) 
\\
&=-\frac{1}{3} \log(z_1 z_2^2) + \biggl(-z_2 -z_3 -6 z_2 z_3 -\frac{3 z_2^2}{2} -\frac{3 z_3^2}{2} +z_1 z_2^2 -30 z_2^2 z_3-30 z_2 z_3^2 
\\
&\qquad-315 z_2^2 z_3^2 +90 z_1 z_2^2 z_3^2+12 z_1 z_2^2 z_3 \biggr)
\\
&\quad
-\left(-z_2 z_3 -10 z_2 z_3^2 -10 z_2^2 z_3 -210 z_2^2 z_3^2 +6 z_1 z_2^2 z_3 +90 z_1 z_2^2 z_3^2\right) \hbar^2
+ \cO(\hbar^4, z_i^3).
\ea\ee
The differential operator giving the leading corrections to the quantum A-periods is given by
\be\ba
 \cD_2 
 =-\frac{z_1}{2}\theta_1 + \left(\frac{1}{12}-z_1 \right) \theta_1^2 +\frac{z_1}{4} \theta_2 + \left(\frac{1}{6} -\frac{z_1}{4} \right) \theta_2^2 + \left(-\frac{1}{12}+z_1 -\frac{1}{12z_1 z_2} \right) \theta_1 \theta_2 .
\ea\ee
Then, we can obtain the quantum mirror maps and quantum B-periods by acting the differential operator on the classical periods,
\be\ba
t^{(2)}_i = \cD_2 t_i,~~ (i=1,2,3),
\qquad
{\Pi}_{d,k}^{(2)} =\cD_2 \Pi_{d,k}^{(0)},
~ ~(k=1,2). 
\label{2ndQuantCorrY31}
\ea\ee

To check the consistency, we calculate the NS free energy near the large radius frame which can be calculated from the general formulae \eqref{F1andF0} and \eqref{F2andF0}.
Then, we find the instanton parts of $\left[ F_n\right]^{\text{inst.}}$ for $n=1,2$,
%or directly solve \eqref{dualPeriod} with \eqref{MatrixY31} recursively, 
%\be\ba
%&\partial_{t_1}F_0 = \frac{1}{3}(2\Pi_{d,1}^{(0)}+\Pi_{d,2}^{(0)} -2 \partial_{t_3} F_0),
%\\
%&\partial_{t_2}F_0 = \frac{1}{3}(\Pi_{d,1}^{(0)}+2\Pi_{d,2}^{(0)} -4 \partial_{t_3} F_0),
%\\
%&\partial_{t_1}F_1 = \frac{1}{3}\biggl\{ 2\Pi_{d,1}^{(2)}+\Pi_{d,2}^{(2)}-2 \partial_{t_3}F_1 -3 \left( t_1^{(2)} \partial_{t_1}^2 F_0 + t_2^{(2)}\partial_{t_1} \partial_{t_2} F_0 + t_3^{(2)}\partial_{t_1} \partial_{t_3} F_0  \right)  
%\\
%&\hspace{59mm}
%-2 \left( t_1^{(2)} \partial_{t_1}\partial_{t_3} F_0 + t_2^{(2)}\partial_{t_2} \partial_{t_3} F_0 + t_3^{(2)}\partial_{t_3}^2 F_0  \right)  
%\biggr\}
%\\
%&\partial_{t_2}F_1 = \frac{1}{3}\biggl\{ \Pi_{d,1}^{(2)}+2\Pi_{d,2}^{(2)}-4 \partial_{t_3}F_1 -3 \left( t_1^{(2)} \partial_{t_1}\partial_{t_2} F_0 + t_2^{(2)}\partial_{t_2}^2 F_0 + t_3^{(2)}\partial_{t_2} \partial_{t_3} F_0  \right)  
%\\
%&\hspace{59mm}
%-4 \left( t_1^{(2)} \partial_{t_1}\partial_{t_3} F_0 + t_2^{(2)}\partial_{t_2} \partial_{t_3} F_0 + t_3^{(2)}\partial_{t_3}^2 F_0  \right)  
%\biggr\}.
%\label{NSY31}
%\ea\ee
%From them, we find the NS free energy whose instanton parts $\left[ F_n\right]^{\text{inst.}}$ are given by
\be\ba
\left[ F_1\right]^{\text{inst.}}
&=\frac{Q_1}{6}+\frac{Q_2}{6}+\frac{Q_3}{6}+\frac{Q_1 Q_2}{6}+\frac{Q_1 Q_3}{6}+\frac{7 Q_2 Q_3}{3}+\frac{5 Q_1 Q_2 Q_3}{2}
 + \cO(Q_i^2),
\\
\left[ F_2\right]^{\text{inst.}}
&=\frac{Q_1}{360}+\frac{Q_2}{360}+\frac{Q_3}{360}+\frac{Q_1 Q_2}{360}+\frac{Q_1 Q_3}{360}-\frac{59 Q_2 Q_3}{180}-\frac{13 Q_1 Q_2 Q_3}{40}+ \cO(Q_i^2).
\ea\ee
They agree with the topological vertex computations.

Now we are ready to calculate the quantum corrections to the energy spectra.
The Bohr--Sommerfeld quantization conditions are given by \eqref{AllOrderQC}, where the quantum corrected spectra and volumes are defined in \eqref{expandEandVol}.
In the classical limit $\hbar =0$ of \eqref{AllOrderQC}, the classical Bohr--Sommerfeld volumes vanish at the classical minimum energies 
\be\ba
E_1^{(0)}= \frac{4-r^9}{r^6} =: E_{\m_1},~~E_2^{(0)}= \frac{2+r^9}{r^{12}} =: E_{\m_2},
\ea\ee 
which correspond to the conifold point.
In the following, we do the computations for a particular case $r=2^{-1/9}$.
The leading corrections to the energy spectra are given by \eqref{1stCorTop}.
By comparing with the perturbative computation \eqref{Y31Spectrum}, we find the exact values of $E_{1,2}$-derivatives of the volumes at the conifold point,
\be{\footnotesize \ba
\begin{pmatrix}
\partial_{E_1} \vol_1^{(0)}  & \partial_{E_2} \vol_1^{(0)} 
\\
\partial_{E_1} \vol_2^{(0)} & \partial_{E_2} \vol_2^{(0)}
\end{pmatrix}
=\pi
\begin{pmatrix}
 -\frac{-2 \sqrt{2}-2 }{2^{2/3} \left(166 \sqrt{2}+245\right)^{1/6}} & \frac{1}{\left(83 \sqrt{2}+\frac{245}{2}\right)^{1/6}} 
 \\
 -\frac{2^{1/3} }{\left( 4 \sqrt{2}+7\right)^{1/2}} & -\frac{2^{11/12} \left(\sqrt{2}+1\right) \left(2^{1/4} \left(166 \sqrt{2}+245\right)^{1/6}-\left( 37 \sqrt{2}-52\right)^{1/2}\right) }{2^{3/4} \left( 7 \sqrt{2}+8\right)^{1/2} \left(166 \sqrt{2}+245\right)^{1/6}-2 \left( 51-34 \sqrt{2}\right)^{1/2}}
\end{pmatrix},
\ea} \ee
which agree with the numerical computations.

In this case, we do not calculate the classical mirror maps around the conifold point, but when one wants to calculate higher corrections to the energy spectra as in the case of $Y^{3,0}$, the classical mirror maps are needed to obtain the higher-order quantum corrections to the (derivatives) of the volumes via the formulae \eqref{volandmap}, \eqref{volandmapDiff}.

%%%%%%%%%%

\subsubsection{$Y^{3,2}$ model}
In this example, we sometimes also use some of the notations and definitions in Section \ref{Y30case}. The mirror curve of $Y^{3,2}$ is
\be\ba
\re^p + z_3 \re^{x-p}+ z_1 z_2^2 \re^{3x} + z_2 \re^{2x}+\re^x +1 =0.
\ea\ee
The Picard--Fuchs operators are
\be\ba
&\cL_1=\theta_1 (\theta_1 -2 \theta_2 - \theta_3) -z_1 (-2\theta_1 + \theta_2 -1)(-2 \theta_1 + \theta_2),
\\
&\cL_2=(\theta_2-\theta_3) (-2\theta_1 + \theta_2) - z_2 (\theta_1 -2 \theta_2 -\theta_3 -1)(\theta_1 -2 \theta_2 -\theta_3),
\\
&\cL_3 =\theta_3^2 - z_3 (\theta_1 - 2\theta_2 -\theta_3)(\theta_2 -\theta_3),
\\
&\cL_4 =\theta_1 \theta_3^2 -z_1 z_2 z_3( \theta_1 -2\theta_2 -\theta_3 -1)(\theta_1 -2\theta_2 -\theta_3)(-2\theta_1 + \theta_2) .
\ea\ee 
The classical mirror maps and the derivatives of the prepotential are given by
\be\ba
t_1({\bm z}) = \omega_1=\log z_1 
&+2 z_{1} -z_{2} +3 z_{1}^2 -\frac{3 z_{2}^2}{2}+2 z_{2} z_{3} -2 z_{1}^2 z_{2} +z_{1} z_{2}^2 +12 z_{2}^2 z_{3} 
\\
&
-15 z_{2}^2 z_{3}^2 -6 z_{1} z_{2}^2 z_{3} +6 z_{1} z_{2}^2 z_{3}^2+ \cO(z_i^3),
\\
t_2({\bm z}) = \omega_2=\log z_2
&-z_{1}+2 z_{2} -\frac{3 z_{1}^2}{2} +3 z_{2}^2 -4 z_{2} z_{3} +z_{1}^2 z_{2} -2 z_{1} z_{2}^2 -24 z_{2}^2 z_{3} 
\\
&+30 z_{2}^2 z_{3}^2 +12 z_{1} z_{2}^2 z_{3} -12 z_{1} z_{2}^2 z_{3}^2 + \cO(z_i^3),
\\
t_3({\bm z}) = \omega_3 =\log z_3
&+\frac{1}{3}\left\{ \left(t_1({\bm z}) -\log z_1 \right) +2 \left(t_2({\bm z}) -\log z_2 \right) \right\},
\ea\ee
and
\be\ba
&\frac{\partial F_0}{\partial t_1} =\frac{1}{18}\left(16 \omega_{11} +10\omega_{12} +\omega_{22} + 12 \omega_{13} +6 \omega_{23} \right),
\\
&\frac{\partial F_0}{\partial t_2} =\frac{1}{18}\left(5 \omega_{11} +2\omega_{12} +2\omega_{22} + 6 \omega_{13} +12 \omega_{23} \right),
\\
&\frac{\partial F_0}{\partial t_3} = \frac{1}{3} \left(\omega_{11} + \omega_{22} + \omega_{12}  \right),
\ea\ee
where $\omega_i$ and $\omega_{ij}$ are defined in \eqref{omega} with the coefficients $c(l,m,n;\rho_i)$,
\be\ba
c(l,m,n;\rho_i) = 
&\frac{\Gamma \left(\rho _1+1\right) \Gamma \left(\rho _3+1\right)^2 \Gamma \left(\rho _2-\rho _3+1\right)}{\Gamma \left(l+\rho _1+1\right) \Gamma \left(n+\rho _3+1\right)^2 \Gamma \left(m-n+\rho _2-\rho _3+1\right) }
\\
&\times 
\frac{ \Gamma \left(-2 \rho _1+\rho _2+1\right) \Gamma \left(\rho _1-2 \rho _2-\rho _3+1\right) }{ \Gamma \left(m-2 \left(l+\rho _1\right)+\rho _2+1\right)\Gamma \left(l-n+\rho _1-2 \left(m+\rho _2\right)-\rho _3+1\right)}.
\ea\ee
The classical B-periods $\Pi_{d,i}~(i=1,2)$ are given by \eqref{PrepotBperiod} with the non-invertible matrix $C_{ij}$,
\be\ba\label{MatrixY32}
C = 
\left[
\begin{array}{rrr}
2 & -1 & 0 \\
-1 & 2 & 1
\end{array}
\right],
\ea\ee
From the prepotential, the Bohr-Sommerfeld volumes are
\begin{eqnarray}
\vol^{(0)}_i({\bm z})=\sum_{i=1}^3C_{ij}\frac{\partial F_0}{\partial t_j} -\frac{2\pi^2}{3},~~~ i=1,2.
\end{eqnarray}  
 The complex structure moduli parameters $z_1, z_2, z_3$ are related to the eigenvalues of the dimer model by 
\begin{eqnarray}
z_1 =\frac{E_2}{E_1^2} , ~~ z_2 =\frac{E_1}{E_2^2}, ~~ z_3 =-\frac{R^6}{E_2}. 
\end{eqnarray} 
From the classical limit $\hbar=0$ of \eqref{Y32Spectrum}, the classical phase volumes vanish at 
\be\ba
z_1 = \frac{r^{9/2}-3r^{9/4}+5}{(2r^{9/4}-5)^2},~~ z_2 = \frac{5r^{9/4}-2r^{9/2}}{(r^{9/2}-3r^{9/4}+5)^2},~~ z_3 = \frac{(r^{9/4}-1)^3}{r^{9/4}(r^{9/2}-3r^{9/4}+5)},
\ea\ee
where we use the polynomial relation \eqref{RandrY32} to eliminate $R$.
We check numerically that the volumes vanish at this point for, e.g., $r=2^{-4/9}$.

Now let us consider the quantum mirror curve,
\be\ba
\Psi(x- \ri \hbar) + \re^{\frac{\ri \hbar}{2}} z_3 \re^{x} \Psi(x+ \ri \hbar) + \left( z_1 z_2^2 \re^{3x} + z_2 \re^{2x}+\re^x +1\right) \Psi(x) =0.
\ea\ee
By taking the residue of $ w(x;\hbar)$, we find a quantum A-period,
\be{\footnotesize \ba
\Pi({\bm z};\hbar) 
&= -\frac{1}{3} \log(z_1 z_2^2) + \oint_{x=\infty} \rd x w(x;\hbar) 
\\
&=-\frac{1}{3} \log(z_1 z_2^2) + \left(-z_{2} -\frac{3 z_{2}^2}{2}+2 z_{2} z_{3} +z_{1} z_{2}^2 +12 z_{2}^2 z_{3}-15 z_{2}^2 z_{3}^2 -6 z_{1} z_{2}^2 z_{3} +6 z_{1} z_{2}^2 z_{3}^2\right)
\\
&\quad
-\left(\frac{z_{2} z_{3}}{4}+\frac{7 z_{2}^2 z_{3}}{2} -\frac{15 z_{2}^2 z_{3}^2}{2}-\frac{11}{4} z_{1} z_{2}^2 z_{3} +5 z_{1} z_{2}^2 z_{3}^2 \right) \hbar^2
+ \cO(\hbar^4, z_i^3).
\ea} \ee
The differential operator giving the leading correction to the quantum A-periods in this case has the following relatively long
expression,
\be\ba
\cD_2
 &= -\frac{1}{24z_2(1-z_3)}\biggl\{
 2(4 -5 z_{3} + 12z_{1} z_{2} -13 z_{1} z_{2}z_{3})\theta_1 
 \\
 &+ \frac{1}{z_{1}} \left(-4+16 z_{1}+5 z_{3} -4 z_{1} z_{2}-20 z_{1} z_{3} +48 z_{1}^2 z_{2} +3 z_{1} z_{2} z_{3} -52 z_{1}^2 z_{2} z_{3}\right) \theta_1^2 
 \\
 &+ (-4 +5 z_{3} -12 z_{1} z_{2} +13 z_{1} z_{2} z_{3}) \theta_2 
 \\
 &+ (4 +8z_{2} -5 z_{3} + 12z_{1}z_{2} -12z_{2} z_{3} -13 z_{1}z_{2} z_{3}) \theta_2^2
 \\
 &+ \frac{1}{z_{1}} \left(12 -16 z_{1}-15 z_{3} +4 z_{1} z_{2}+20 z_{1} z_{3} -48 z_{1}^2 z_{2} + 52 z_{1}^2 z_{2} z_{3}\right) \theta_1 \theta_2  
 \biggr\} .
\ea\ee
Then, we can obtain the quantum mirror maps and quantum B-periods by acting above operator on the classical periods, as in \eqref{2ndQuantCorrY31}. 

We do not provide the details of the calculations of the NS free energy in this case since the computation process is completely the same as the $Y^{3,1}$ model, but one can show that the NS free energy calculated from the differential operators agree with the topological vertex computations.

Now we are ready to calculate the quantum corrections to the energy spectra.
The Bohr--Sommerfeld quantization conditions are given by \eqref{AllOrderQC}, where the quantum corrected spectra and volumes are defined in \eqref{expandEandVol}.
From the classical limit $\hbar=0$ of \eqref{AllOrderQC}, the classical Bohr--Sommerfeld volumes vanish at the classical minimum energies,
\be\ba
E_1^{(0)}= \frac{5-r^{9/4}}{r^{3/4}} =: E_{\m_1},~~E_2^{(0)}= \frac{5}{r^{3/2}} -3 r^{3/4} + r^3 =: E_{\m_2}, 
\ea\ee 
which correspond to the conifold point.
For simplicity, we do the computations for $r=2^{-4/9}$.
The leading corrections to the spectra are given by \eqref{1stCorTop}.
By comparing with the perturbative computations, we find the exact values of $E_{1,2}$-derivatives of the volumes at the conifold point,
\be \ba
\begin{pmatrix}
\partial_{E_1} \vol_1^{(0)} &\partial_{E_2} \vol_1^{(0)}
\\
\partial_{E_1} \vol_2^{(0)} & \partial_{E_2} \vol_2^{(0)} 
\end{pmatrix}
=\frac{\pi}{3}
\begin{pmatrix}
\frac{2^{11/3} }{ \sqrt{7}} &  \frac{2^{7/3} }{ \sqrt{7}}
\\
 -2^{2/3}  & -2^{4/3}
\end{pmatrix},
\ea \ee
which are consistent with the numerical computations.

Similar to the $Y^{3,1}$ case, we do not calculate the classical mirror maps around the conifold point, but when one calculates higher-order corrections to the energy spectra as in the $Y^{3,0}$ model, the classical mirror maps are needed to obtain the higher-order quantum corrections to the (derivatives) of the volumes via the formulae \eqref{volandmap} \eqref{volandmapDiff}.

%%%%%%%%%%

\subsubsection{$Y^{3,3}$ model}
As the final example, we consider the $Y^{3,3}$ model.
We sometimes use some of the notations and definitions in Section \ref{Y30case}.
The mirror curve of $Y^{3,3}$ is
\be\ba
 \re^{p} + \re^{-p} + \re^{3x} + \frac{\re^{2x} }{z_1^{1/3} z_2^{2/3} z_3^{1/6}}+ \frac{ \re^{x}}{z_1^{2/3} z_2^{1/3} z_3^{1/3}} + \frac{1}{z_3^{1/2}} =0.
\ea\ee
The Picard--Fuchs operators are
\be\ba
&\cL_1=\theta_1 \left(\theta_1 -2\theta_2 -2\theta_3 \right) + 4 \theta_2 \theta_3 -z_1 \left(2\theta_1 -\theta_2 +1 \right)\left( 2\theta_1 -\theta_2\right),
\\
&\cL_2=\theta_2\left(\theta_2 -2\theta_1 \right) +z_2 \left(2\theta_2 -\theta_1 +1 \right)\left( 2\theta_2 -\theta_1\right),
\\
&\cL_3=\theta_3^2 +z_3 \left(2\theta_3 -\theta_1 +1 \right)\left( 2\theta_3 -\theta_1\right),
\\
&\cL_4 = \theta_2 \theta_3^2 + z_1 z_2 z_3 (\theta_1-2\theta_2)(\theta_1-3\theta_3)(2\theta_1-\theta_2).
\ea\ee
Their solutions provide the mirror maps and the derivatives of prepotential,
\be {\footnotesize \ba
&t_1({\bm z}) = \omega_1=\log z_1 +2 z_1 -z_2 -z_3 +3 z_1^2 -\frac{3 z_2^2}{2}-\frac{3 z_3^2}{2} -2 z_1^2 z_2+6 z_1^2 z_3+z_1 z_2^2 -4 z_1^2 z_2 z_3+ \cO(z_i^3),
\\
&t_2({\bm z}) = \omega_2=\log z_2 -z_1 +2 z_2 -\frac{3 z_1^2}{2} +3 z_2^2 +z_1^2 z_2-3 z_1^2 z_3 -2 z_1 z_2^2 + 2 z_1^2 z_2 z_3+ \cO(z_i^3), 
\\
&t_3({\bm z}) = \omega_3 =\log (z_3)+2 z_3 +3 z_3^2 + \cO(z_i^3),
\ea } \ee
and
\be\ba
&\frac{\partial F_0}{\partial t_1} =\frac{2}{3} \omega_{11} +\frac{2}{3} \omega_{12} +\frac{2}{3} \omega_{13} +\frac{2}{3}\omega_{22} + \frac{1}{3} \omega_{23} +\frac{2\pi^2}{3},
\\
&\frac{\partial F_0}{\partial t_2} =\frac{1}{3} \omega_{11} +\frac{4}{3} \omega_{12} +\frac{1}{3} \omega_{13} +\frac{4}{3}\omega_{22} + \frac{2}{3} \omega_{23} +\frac{2\pi^2}{3},
\ea\ee
where $\omega_i$ and $\omega_{ij}$ are given in \eqref{omega} with the coefficients $c(l,m,n;\rho_i)$,
\be\ba
c(l,m,n;\rho_i) = 
&\frac{1}{\Gamma(1+l-2m+\rho_1-2\rho_2)^2 \Gamma(1-2l + m - 2\rho_1 + \rho_2) \Gamma(1+l-2n+\rho_1-2\rho_3)} 
\\
&\times \frac{1}{ \Gamma(1+m+\rho_2)\Gamma(1+n+\rho_3)^2}
.
\ea\ee
For the third mirror map $t_3$, the summation can be expressed in a closed form,
\be\ba
t_3({\bm z}) = \log z_3 - 2 \log \left(\frac{1-\sqrt{1-4 z_3}}{2} \right).
\ea\ee
The classical B-periods are completely the same form as the ones of $Y^{3,0}$ since the matrices $C_{ij}$ of $Y^{3,0}$ and $Y^{3,3}$ are the same.
From the prepotential, the Bohr-Sommerfeld volumes are
\begin{eqnarray}
\vol^{(0)}_i({\bm z})=\sum_{j=1}^3C_{ij}\frac{\partial F_0}{\partial t_j} -\frac{2\pi^2}{3} ,~~~ i=1,2, 
\end{eqnarray}  
where the complex structure moduli parameters $z_1, z_2, z_3$ are related to the eigenvalues of dimer model by 
\begin{eqnarray}
z_1 =\frac{(1+R^6)E_1}{E_2^2} , ~~ z_2 =\frac{E_2}{E_1^2}, ~~ z_3 =\frac{R^6}{(1+R^6)^2}. 
\end{eqnarray} 
The Bohr-Sommerfeld volumes should vanish at the conifold point,
\be\ba
z_1 = \frac{(1+R^2)(1+R^6)}{3(1+R^2+R^4)^2},
\qquad
z_2 = \frac{1+R^2+R^4}{3(1+R^2)^2},
\qquad
z_3 = -\frac{R^6}{(1+R^6)^2}.
\ea\ee
We check that the volumes vanish numerically for, e.g., $R=1$.

Now let us consider the quantum mirror curve given by
\be\ba
\Psi(x+i \hbar)+ \Psi(x-i \hbar)+\left( \re^{3x} +\frac{\re^{2x}}{z_1^{1/3}z_2^{2/3}z_3^{1/6}}+\frac{\re^x}{ z_1^{2/3}z_2^{1/3}z_3^{1/3}}+ \frac{1}{z_3^{1/2}}\right)\Psi(x) =0.
\ea\ee
By taking the residue of $ w(x;\hbar)$, we find a quantum A-period,
\be\ba
\Pi({\bm z};\hbar) 
&= -\frac{1}{3} \log(z_1^2 z_2z_3) + \oint_{x=\infty} \rd x w(x;\hbar) 
\\
&= -\frac{1}{3} \log(z_1^2 z_2z_3) + \left(2 z_1^2 z_2 z_3+z_1^2 z_2-3 z_1^2 z_3-\frac{3 z_1^2}{2}-z_1 \right)
\\
&\qquad-\left(z_1^2 z_3 -z_1^2 z_2 z_3\right) \hbar^2
+ \cO(\hbar^4, z_i^3).
\ea\ee
The differential operator giving the leading correction to the quantum A-periods is given by
\be\ba
 \cD_2 
&=\frac{1}{12(-1+2z_2)}\biggl\{
 \frac{z_1 \left(15 z_2^2-12 z_2+4\right)-z_2}{2} \theta_1 
 + \frac{z_1 \left(9 z_2^2-24 z_2+8\right)+5 z_2-2}{2} \theta_1^2
 \\
&
+(z_1 \left(-15 z_2^2+6 z_2-1\right)+z_2) \theta_2
 -\frac{z_1 \left(36 z_2^2-15 z_2+2\right)-4 z_2+1}{2} \theta_2^2
 \biggr\}.
\ea\ee
Then, we can obtain the quantum B-periods by acting above differential operator on the classical B-periods.

We do not provide the details of the calculations of the NS free energy in this case since the computation process is completely the same as $Y^{3,0}$ model, but one can show that the NS free energy calculated from the differential operators agrees with the topological vertex computations.

Now we are ready to calculate the quantum corrections to the energy spectra. 
For simplicity, we do the computations for $R=1$.
The Bohr--Sommerfeld quantization conditions are given by \eqref{AllOrderQC}, where the quantum corrected spectra and volumes are defined in \eqref{expandEandVol}.

In the classical limit $\hbar =0$ of \eqref{AllOrderQC}, the classical Bohr--Sommerfeld volumes vanish at the classical minimum energies
\be\ba
E_1^{(0)}= 6 =: E_{\m_1},
\qquad
E_2^{(0)}= 9 =: E_{\m_2},
\ea\ee 
which correspond to the conifold point. 
The leading corrections to the spectra are given by \eqref{1stCorTop}.
By comparing them with direct perturbative calculations (\ref{Y33Spectrum}), we find the exact values of $E_{1,2}$-derivatives of the volumes at conifold point,
\be\ba
\begin{pmatrix}
\partial_{E_1} \vol_{1}^{(0)} & \partial_{E_2} \vol_{1}^{(0)}
\\
\partial_{E_1} \vol_{2}^{(0)} & \partial_{E_2} \vol_{2}^{(0)} 
 \end{pmatrix} 
=
\begin{pmatrix}
\sqrt{3} & -\frac{1}{\sqrt{3}}
\\
-\frac{1}{\sqrt{3}} & \frac{1}{\sqrt{3}} 
 \end{pmatrix}.
\ea\ee
With the change of variables, we find
\begin{eqnarray}
\begin{pmatrix}
 \partial_{z_1}\vol^{(0)}_1({\bm z}) & \partial_{z_2}\vol^{(0)}_1({\bm z}) \\ 
  \partial_{z_1}\vol^{(0)}_2({\bm z}) & \partial_{z_2}\vol^{(0)}_2({\bm z})  
 \end{pmatrix} 
 =
\sqrt{3} \pi \begin{pmatrix}
 -9 & \frac{4}{3} \\ 
  0 & -12 
 \end{pmatrix} .
\end{eqnarray}
We check numerically that this is indeed true.

The classical A-periods near the conifold point are
\be {\footnotesize \ba
t_{c,1} 
=&-3 \pi \sqrt{3} z_{c,1} -\frac{68 \pi z_{c,2}}{3 \sqrt{3}} -\frac{20 \pi z_{c,3}}{9 \sqrt{3}}+\frac{131 \pi z_{c,1} z_{c,2}}{2 \sqrt{3}}+\frac{7 \pi z_{c,1} z_{c,3}}{2 \sqrt{3}}+\frac{400 \pi z_{c,2} z_{c,3}}{27 \sqrt{3}}  -\frac{1657 \pi z_{c,1} z_{c,2} z_{c,3}}{36 \sqrt{3}}
\\
&+ \cO(z_{c,i}^2),
\\
t_{c,2} 
=&-6 \pi \sqrt{3} z_{c,1} -\frac{28 \pi z_{c,2}}{3 \sqrt{3}}-\frac{28 \pi z_{c,3}}{9 \sqrt{3}} +\frac{19 \pi z_{c,1} z_{c,2}}{2 \sqrt{3}}+\frac{23 \pi z_{c,1} z_{c,3}}{2 \sqrt{3}}+\frac{80 \pi z_{c,2} z_{c,3}}{27 \sqrt{3}} + \frac{7 \pi z_{c,1} z_{c,2} z_{c,3}}{36 \sqrt{3}}
\\
&+ \cO(z_{c,i}^2),
\\
t_{c,3} =&-2 \log(1-2\sqrt{-z_{c,3}}) + \log(1+4 z_{c,3}),
\ea } \ee
where
\be\ba
z_1 = \frac{1}{6} + z_{c,1},\quad
z_2 = \frac{1}{6} + z_{c,2},\quad
z_3 = -1 + z_{c,3}.
\ea\ee
The coefficients of $z_{c,1}$ and $z_{c,2}$ are fixed by the relation \eqref{volandmapDiff}.

Repeating the computations for $\hbar^2$-order, we find
\be\ba
&E^{(2)}_{1} =\frac{1}{36} (9 n_1 (n_1+1)+3 n_2 (n_2+1)+8),
\\
&E^{(2)}_{2} =\frac{1}{2}n_1 +n_2^2+\frac{3 n_2}{2} + n_1 n_2 +\frac{2}{3}.
\ea\ee
These results agree with \eqref{Y33Spectrum} for $R=1$.

%%%%%%%%%%%%%%%%%%%%%%%%%%%%%%%%%%%%%%%%%%%%%%%%%

\section{Discussions}

In this paper, we studied the analytic connections between genus two mirror curves and $Y^{p,q}$ cluster integrable systems, which are generalizations of affine $A$-type relativistic Toda systems. It is interesting to consider the much higher genus mirror curves and the application to other types of affine Toda systems. 

In the topic of the differential operator method, there are still interesting issues to clarify.
For example, it would be interesting to consider the genus one mirror curves for local $E_n$ del Pezzo surfaces, where the global symmetries are $E_n$ groups. Such curves are considered in \cite{HKRS, HKP} with some mass parameters turned off. With all mass parameters turned on where the Calabi-Yau threefolds are non-toric, it is interesting to study the differential operator approach for the cases. %\cite{HSW}.

Also, in \cite{FMS}, the authors pointed out that the quantum A-periods of $D_5$ del Pezzo geometry can be expressed as $D_5$ Weyl characters.
The quantum mirror map of this curve would be also given in the differential operator method.
Therefore, it would be interesting to clarify relations between the Weyl group expression and the differential operators.

Recently, the authors in \cite{AGH} provide the analytic results on the black hole perturbation theory from the quantization conditions. They consider the quantization conditions for A-periods, not B-periods.
It would be interesting to clarify the physical implications of this quantization conditions in the integrable systems or topological strings.

\subsubsection*{Acknowledgements}
We would like to thank Sheldon Katz, Albrecht Klemm for helpful discussions and/or stimulating collaborations on related papers.
Some of the computations based on {\it mathematica} were carried out on the computer {\it sushiki} at Yukawa Institute for Theoretical Physics at Kyoto University.
The work of MH and YS was supported by the national Natural Science Foundation of China (Grants No.11675167 and No.11947301).

%%%%%%%%%%%%%%%%%%%%%%%%%%%%%%%%%%%%%%%%%%%
\appendix

\section{An eigenvalue formula} \label{AppendixEigenvalue} 

Suppose $S$ is a real symmetric $2n\times 2n$ matrix, and $M$ is a real symplectic $2n\times 2n$ matrix that diagonalizes the symmetric matrix, i.e., we have 
\begin{eqnarray}
\Sigma = \begin{pmatrix}
 0 & I_n \\ 
 -I_n & 0  
 \end{pmatrix} ,~~~ M^T \Sigma M =\Sigma, ~~~ M^T S M = \begin{pmatrix}
 C & 0 \\ 
 0 & D  
 \end{pmatrix} ,
\end{eqnarray}  
where $C=\textrm{diag}\{ c_1, c_2, \cdots , c_n \}, D=\textrm{diag}\{ d_1, d_2, \cdots , d_n \} $ are real $n\times n$ diagonal matrices. Then we can show that the characteristic polynomial of the matrix $S\Sigma$ (or $\Sigma S$) is 
\begin{eqnarray} \label{characterB2} 
\det (S\Sigma -\lambda I) = \prod_{k=1}^n (\lambda^2 +c_k d_k) . 
\end{eqnarray}  
So the eigenvalues of $S\Sigma$ are $\pm i \sqrt{c_kd_k}, k=1,2, \cdots n$. In the context of our physics problem, the two diagonal matrices are identical $C=D$, therefore the diagonal elements are completely determined by the symmetric matrix $S$, are thus independent of the choice of the symplectic matrix $M$. 

The calculations are straightforward. Noticing $\Sigma^2 =- I$ and $(-\Sigma M^T) (\Sigma M) =I$, so the characteristic polynomial is 
\begin{eqnarray}
\det (S\Sigma -\lambda I) &=& \det (-\Sigma M^T S\Sigma^2 M -\lambda I) \nonumber \\ &=&
 \det ( \begin{pmatrix}
 0 & D \\ 
 -C & 0  \end{pmatrix} -\lambda I) . 
\end{eqnarray}  
It is now simple to verify the determinant is indeed the polynomial in the right hand side of (\ref{characterB2}).

\addcontentsline{toc}{section}{References}

\providecommand{\href}[2]{#2}\begingroup\raggedright\endgroup

%\bibliographystyle{../bib/utphys}

%\bibliographystyle{utphys} 
%\bibliography{Ref_all}

%\addcontentsline{toc}{section}{References}
%\begin{thebibliography}{10}
    
%\end{thebibliography}
\end{document}